\documentclass[review,3p,11pt]{elsarticle}

\pdfoutput=1

\usepackage{amsfonts, amssymb, amsmath, amscd}
\usepackage{enumerate}
\usepackage{graphicx}
\usepackage[dvips]{epsfig}
\usepackage{color}
\usepackage{booktabs}
\usepackage{soul}
\usepackage{todonotes}

\usepackage{subfig}

\graphicspath{{figs/}}

\newcommand{\mcal}[1]{\mathcal{#1}}

\journal{}

\begin{document}


\begin{frontmatter}



\title{Validating Predictions of Unobserved Quantities}


\author[pecos]{Todd A.~Oliver\corref{oliver}}
\ead{oliver@ices.utexas.edu}

\author[usc]{Gabriel Terejanu}
\ead{terejanu@cec.sc.edu}

\author[pecos]{Christopher S.~Simmons}
\ead{csim@ices.utexas.edu}

\author[pecos,meche]{Robert D.~Moser}
\ead{rmoser@ices.utexas.edu}

\address[pecos]{Center for Predictive Engineering and Computational Sciences,\\ 
Institute for Computational Engineering and Sciences, 
The University of Texas at Austin}

\address[usc]{Department of Computer Science and Engineering, 
University of South Carolina}

\address[meche]{Department of Mechanical Engineering, 
The University of Texas at Austin}

\cortext[oliver]{Corresponding author: 201 E. 24th Street, C0200, Austin, Texas 78712, USA.}

\begin{abstract}
The ultimate purpose of most computational models is to make
predictions, commonly in support of some decision-making process
(e.g., for design or operation of some system). The quantities that
need to be predicted (the quantities of interest or QoIs) are
generally not experimentally observable before the prediction, since
otherwise no prediction would be needed.  Assessing the validity of
such extrapolative predictions, which is critical to informed
decision-making, is challenging.  In classical approaches to
validation, model outputs for observed quantities are compared to
observations to determine if they are consistent.  By itself, this
consistency only ensures that the model can predict the observed
quantities under the conditions of the observations.  This limitation
dramatically reduces the utility of the validation effort for decision
making because it implies nothing about predictions of unobserved QoIs
or for scenarios outside of the range of observations.  However, there
is no agreement in the scientific community today regarding best
practices for validation of extrapolative predictions made using
computational models.  The purpose of this paper is to propose and
explore a validation and predictive assessment process that supports
extrapolative predictions for models with known sources of error.  The
process includes stochastic modeling, calibration, validation, and
predictive assessment phases where representations of known sources of
uncertainty and error are built, informed, and tested.  The proposed
methodology is applied to an illustrative extrapolation problem
involving a misspecified nonlinear oscillator.
\end{abstract}


\end{frontmatter}


%
\section{Introduction and Motivation} 
\label{sec:introduction}
Advances in computing hardware and algorithms in recent decades, along
with accompanying advances in the fidelity of computational models,
enable simulation of physical phenomena and systems of unprecedented
complexity.  This capability is revolutionizing engineering and science.
For example, results of computational simulations are used heavily in
the design of nearly all complex engineering systems, from consumer
electronics to spacecraft and nuclear power plants.  Furthermore,
results of computational models are used to inform policy decisions in
areas where the consequences of inaccurate predictions and
poorly-informed decisions could be catastrophic, such as disaster
response and climate change. However, for computational modeling to
realize its full potential in such applications, it is critical that the
reliability of the results of the models be systematically
characterized. In the current paper, an approach to making such
reliability assessments is proposed for a broad class of problems in
computational science and engineering.

Any reliability assessment of a computational model must necessarily
consider the purpose for which the model is to be used. A common use of
computational models is to predict the response of some complex system
and thereby to inform some decision regarding the system. For example,
predictions from computational models might be used to decide which of
several competing system designs will be superior, to decide whether a
proposed use scenario for a system will meet operational objectives, or
to decide how to respond to a system as it evolves. Typically, the
decisions are informed by predictions of specific quantities describing
the response of the system, these are the so-called quantities of
interest (QoIs). An important feature of such uses of computational
models is that observational data for the QoIs in the prediction
scenario are not available, since otherwise the predictions would not be
needed. This is the situation of interest here.
There are of course other modes in which computational
models are used in science and engineering, and in those cases the
reliability issues will be different.

The assessment of reliability of computational models is often divided
into three aspects: verification, validation and uncertainty
quantification (collectively known as V\&V-UQ).  Verification is
concerned with assessing the discrepancy between the computer simulation
and the underlying mathematical model on which it is based.  Uncertainty
quantification is the process of assessing uncertainties that affect
simulation predictions, such as those due to uncertain model inputs or
inaccuracies in the model itself, and determining the resulting
uncertainty in the QoIs.  Finally, following~\cite{AIAA_Guide_1998,
ASME_VV10_2006, ASME_VV20_2009}, validation is the process of
determining whether a mathematical model is a sufficient representation
of reality for the purposes for which the model will be used; that is,
for predicting specified QoIs to inform a specific decision.  Thus,
while verification is a purely mathematical process concerned only with
the difference between computational and mathematical models, UQ and
validation are concerned with the discrepancy between the mathematical
model and the real world.

While verification is vitally important and sometimes overlooked in
practice, it is largely understood~\cite{Roache_2009, Oberkampf_2010},
with well developed techniques for estimating and controlling the impact
of numerical errors on specified QoIs \cite{Ainsworth2000,Prudhomme1999}. 
In the remainder of this
paper, it will be assumed that numerical solutions have been verified to
ensure that numerical errors in the predictions are small compared to
other sources of uncertainty, and so, verification will not be discussed
further. Instead, we focus on the validation of and quantification of
uncertainty in predictions of unobserved QoIs.

\subsection{Validation Processes}
In engineering practice, the ``validity'' of a computational model is
often assessed by simply comparing the output of the model with
experimental data.  While such comparisons must be a part of any
validity assessment, a straightforward process considering only the
closeness of this comparison has two important shortcomings when used
to assess the reliability of model predictions. First, it does not
account for the uncertainties associated with the model or the
data. Second, it precludes an assessment of the reliability of
predictions in new situations or for unobserved quantities,
which as discussed above, is the use we consider here. In essence,
predictions are always extrapolations from available information, and
the validation question is whether such extrapolation is justified.

In this context then, it is the validity of the prediction that is
really of interest. We cannot speak of the validity of a model in
general, since a model may produce valid predictions of some quantities
in some circumstances, and not of others. This is different from the
notion of validity of scientific theories, in which we insist that any
inconsistency between a theory and observations falsifies the
theory. In computational modeling, models known to be false in this strict
scientific sense are used routinely (e.g., Newtonian mechanics), and the
resulting predictions can nonetheless be valid, provided the
inadequacies of the model have no significant effect on the prediction.

To address the shortcomings of naive comparisons with data as a
technique for validating extrapolative predictions, a number of more
sophisticated procedures have been proposed~\cite{Sargent_1998,
Balci_1997, ObTr07, BaNoTe08, Roache_2008, Roache_2009,
Oberkampf_2010, Bayarri2007, Higdon2005, Craig2001,
MorrisonJ_CMA_2013}, and validation guidelines have been developed by
professional engineering societies~\cite{AIAA_Guide_1998,
ASME_VV10_2006, ASME_VV20_2009}. These have generally been positive
developments, but they also have shortcomings. Most commonly, they do
not directly address the validity of models to make predictions of
unobserved QoIs.  For instance, Higdon et al.~\cite{Higdon2005} and
Bayarri el al.~\cite{Bayarri2007} present similar validation
frameworks for computational models, based on the work of Kennedy and
O'Hagan on model calibration and model discrepancy
representations~\cite{Kennedy2001}.  These frameworks rely on
statistical models---specifically Gaussian process models---to
represent the difference between the model outputs and observational
data.
However, such representations are insufficient for validation of the
ability to predict unobserved QoIs, because the discrepancy model is
posed only for the observable quantities and there is generally no
direct mapping from the observables to the QoIs.  

A different approach in addressing the inadequacies in model structure
is given by Strong et al.~\cite{Strong_2011}. The main idea is to
decompose the model into sub-functions and make judgments about the
discrepancy of each sub-function. Strong et al.~\cite{Strong_2011}
propose a model refinement strategy based on sensitivity analysis to
quantify the relative importance of different structural errors within
a health economics model.  The strategy of introducing discrepancy
terms within the model at the source of the structural error is also
embraced in the present paper.  We leverage this technique to
create a direct mapping between the discrepancy terms and both the
observables and the unobserved QoIs.  However, a number of challenges
need to be addressed in using this approach for extrapolative
predictions. First, since the discrepancy representation is a
statistical model, it is highly dependent on calibration against
observations and, hence, should not be used in situations in which it
cannot be trained and tested. Thus, in general, use of this sort of
discrepancy model in extrapolative predictions is suspect.  Unlike
Strong et al.~\cite{Strong_2011} which use this strategy to guide
model refinements based solely on building discrepancy models on prior
knowledge, for our physics-based extrapolation problem we introduce a
systematic calibration, validation and predictive assessment of these
discrepancy terms to build confidence in their predictive capability.
Furthermore, since there is no unique decomposition of the model,
identifying the sources of error remains an issue.  In our approach,
we leverage the fact that the models are physics-based.  Such models
are generally constructed from highly reliable physical theories
coupled with less reliable models to close the governing equations.
This structure gives us a unique view of the modeling error.  Since
modeling errors are introduced entirely through the empirical models,
we only need to introduce discrepancy terms where these
embedded empirical models enter the formulation.

In pioneering work by Babu\v{s}ka et al~\cite{BaNoTe08}, the impact of
model discrepancy on unobserved QoIs was accounted for in the
validation process. They addressed a structural mechanics problem in
which the primary modeling challenge was the statistical
representation of the unknown, spatially varying elastic modulus. In
this case, it was expected that the statistical model could be
calibrated to be consistent with observations from each of several
available experimental scenarios, but that these separate calibrations
need not be consistent with each other. The validation question in
this case, was whether any such calibration discrepancies were
important to the prediction of the QoI. This was tested by comparing
the predictions arising from the different calibrations. This approach
can be generalized to other problems in which the model
parameterization is sufficiently rich that the model can always be
adjusted to fit data from a single experimental
scenario~\cite{Oden2013,MorrisonJ_CMA_2013,Panesi2012}.  While this
was true for the problem and models investigated by Babu\v{s}ka et al,
with many engineering models this is not the case. Indeed, a common
symptom of model inadequacy is that a model cannot be calibrated to
match experimental data within experimental uncertainty, even for a
single experimental scenario. In such cases, the Babu\v{s}ka et al
validation approach is not able to account for the impact of these
discrepancies on the QoIs, and thus, a more general formulation is
needed.


\subsection{Predictive Validation}
As discussed above, there are currently no established techniques for
assessing the validity of predictions of unobserved QoIs. Here we
address this problem by proposing a process we call ``predictive
validation'' by which it is possible to test the validity of such
extrapolative predictions in a broad class of problems. In defining
predictive validation, we will also specify the model characteristics
that enable reliable predictions, a set of necessary conditions for
extrapolative predictions and processes that allow satisfaction of these
necessary conditions to be tested.

The predictive validation process described here was developed to
evaluate predictions regarding physical systems.  Such systems are
commonly described by models based on highly reliable theory (e.g.,
conservation laws), whose validity is not in question in the context of
the predictions to be made. However, these highly reliable theories
typically must be augmented with one or more ``embedded models,'' which
are less reliable. The less reliable embedded models may embody various
modeling approximations, empirical correlations, or even direct
interpolation of data.  For example, in continuum mechanics, the
embedded models might include constitutive models and boundary
conditions, while in molecular dynamics they would include models for
interatomic potentials.  We will refer to such models---i.e.,
highly-fidelity models with lower-fidelity embedded components---as
composite models.

The fact that the composite models used for prediction are built on
highly reliable models is an important ingredient enabling reliable
extrapolation in the approach described here, even though less
reliable embedded models are also used.  Specifically, we require that
the less reliable models are not used outside the range where
they have been calibrated and tested.  This restriction does not
necessarily limit our ability to extrapolate using the composite model
since the relevant scenario space for each embedded model is specific
to that embedded model, not the composite model in which it is
embedded.

Another important ingredient is a representation of the uncertainty.
We require mathematical representations of existing or prior
information, the uncertainty in the observational data that will be
used to inform and test the model, and the uncertainty in model
predictions resulting from model inadequacy.  Since Bayesian
probability~\cite{Cox_1961,Jaynes_2003,VanHorn} provides a powerful
representation of uncertainty, probabilistic models are used to
describe uncertainty in this work, and the development of these
probabilistic models is the first step of the predictive validation
process.  In particular, as in previous work~\cite{Kennedy2001,
Higdon2005, Bayarri2007}, our approach relies heavily on statistical
models of model discrepancy.  However, unlike previous work, we take
advantage of the structure of the composite model described above to
introduce discrepancy models that can be used to evaluate uncertainty
in unobserved quantities.  This advance is accomplished by posing a
model for the error in the embedded physical model directly, rather
than a model for the discrepancy between observations and the model output for the
observables.

Prior information on parameters and experimental uncertainty is also
represented using Bayesian probabilistic models.  In this way, we
strive to account for all significant sources of uncertainty,
including uncertainty in model inputs (e.g., parameters and boundary
conditions), model errors, and observational data; so that the
uncertainties in any model output, whether observed or unobserved, are represented.  The
original composite model plus these uncertainty models provides a
complete model representing both our knowledge of the physics as well
as important sources of uncertainty.  This complete model is then the
subject of the predictive validation process.

Typically, both the embedded physical models and the accompanying
uncertainty models have parameters that must be determined from
observations through calibration. In our approach, calibration is the
first in an integrated, three-step process for assessing the
reliability of extrapolative predictions issued by the complete model
described above.  The three steps---calibration, validation, and
predictive assessment---are designed to answer three distinct
questions regarding the model and its predictive capability. In
calibration, the model is informed by data.  Specifically, parameter
values and their uncertainties are inferred from available
observations by solving an inverse problem.  The use of probability to
represent uncertainty naturally leads to the formulation of the
calibration problem as a Bayesian update.

In validation, outputs from the calibrated model are checked for
consistency with available observations.  While this consistency is
not sufficient for predictive validation, it is necessary.  Unlike
calibration, the validation step is not naturally expressed in the
language of Bayesian hypothesis testing because there are not
well-defined alternative hypotheses.  
%
Instead, we must assess whether the validation data are plausible
according to the model.  There are a number of possibilities for
quantifying this plausibility.  Here we use highest posterior density
credible sets.

Finally, predictive assessment determines whether the calibration and
validation phases were sufficiently informative and challenging to
provide confidence in the reliability of the predictions of the QoIs.
Specifically, one must answer a number of questions about what the
validation tests imply about the QoI predictions.  This assessment
relies on sensitivity analysis along with knowledge about
the embedded models and their formulation.

Notice that in this approach to predictive validation, calibration and
validation are based primarily on statistical analysis.  However, all
representations of uncertainty (e.g., models of uncertainty due to
model discrepancy) depend heavily on the structure of the physical
model and knowledge about the physical system being modeled.  This
knowledge is also crucial to the predictive assessment.  It is
this reliance on knowledge of and reliable theory about
the physical system being modeled that makes extrapolative prediction
possible.

The remainder of the paper is organized as follows.  Important
features of a composite model and associated uncertainty representations
that enable predictive validation for unobserved QoIs are described
abstractly in~\S\ref{sec:overview}.   Specific procedures used to assess the model and build
confidence in its predictive capability is introduced in \S\ref{sec:predval}.  The main ideas and tools of
the predictive validation approach are illustrated using a simple
example involving extrapolative predictions made for a
spring-mass-damper system in~\S\ref{sec:results}.  Concluding remarks
in~\S\ref{sec:conclusions} address the practical challenges in carrying
out the proposed validation processes for real-world applications.

%
\section{Model Inadequacy and Discrepancy Representations in Predictive Validation}
\label{sec:overview}
In developing the predictive validation process to be described, it is
important to be precise about what constitutes a prediction. Here, a
prediction is the result of a computational simulation conducted to
compute specific QoIs which are to be used to support some decision
regarding the system being simulated.  Further, there is no
observational data available for the QoIs for the scenarios of interest,
since otherwise predictions would not be necessary.  Thus, the
credibility of the prediction must established based on available data
for other quantities and/or scenarios, along with any other available
information about the system. The fundamental challenge is to make
credible predictions, despite the fact that the predictions are
extrapolations from available information. Furthermore, in recognition
of the fact that observational data have uncertainties, and that models
are imperfect, we insist that predictions must be endowed with
characterized uncertainties.

In general, the need to extrapolate raises concerns about the
reliability of the predictions, and, indeed, we must ask: ``what
entitles us to make such predictions?'' Part of the answer is that,
unlike purely empirical models that can be used solely to represent
observations, the models used to predict the behavior of physical
systems are often based on theories that are known to be highly reliable
within well-defined domains of applicability.  However, these highly
reliable theories are usually augmented with one or more ``embedded
models,'' which are less reliable.  Thus, the model as a whole is a
composite model, as defined in~\S\ref{sec:introduction}.  In the
predictive validation approach proposed here, reliable predictions are
enabled by the reliable theories that form the foundation of the composite model,
whose validity in the prediction scenario is not in doubt. 

To make these ideas clear, a simple abstract prediction problem is
presented in \S\ref{sec:problem}. This is the simplest problem that has
the characteristics of prediction discussed above. Critical to the
prediction validation process is the representation of uncertainty due
to the imperfections of the model. Background on such discrepancy
representations is provided in
\S\ref{sec:back_discrepancy}, and discrepancy modeling for predictive
validation is described in \S\ref{sec:modeling}. Finally,
generalizations of the abstract problem described in \S\ref{sec:problem}
to encompass predictions in complex physical systems are discussed in
\S\ref{sec:genproblem}. 

\subsection{Abstract Problem Statement} \label{sec:problem}
Reliable predictions are enabled by the use of reliable physical theory
whose validity in the context of the predictions to be made is not in
doubt. Let this theory be written mathematically as
\begin{equation}
\label{eqn:system}
\mathcal{R}(u,\tau;r) = 0,
\end{equation}
where $\mathcal{R}$ is an operator expressing the theory. For example,
in continuum mechanics, $\mathcal{R}$ would be a partial differential
operator expressing conservation of mass, momentum, and energy. In this
formulation, $u$ is the solution or state variable, and $r$ is a set of
scenario variables needed to precisely define the problem being
considered. The scenario variables might include the geometry of the
solution domain, boundary conditions, and other parameters that define
the problem. The final variable $\tau$ is a quantity that needs to be
known to solve~\eqref{eqn:system}. For example, in continuum
mechanics, $\tau$ could be the strain energy or the stress tensor.

If $\tau$ were known in terms of $u$ and $r$, the system would be
closed, and~\eqref{eqn:system} would implicitly define a mapping from
the scenario variables $r$ to the solution variables $u$.  However,
it is often the case that the required relationship between $\tau$ and
$u$ and $r$ is unknown or does not exist---i.e., $u$ and $r$ do
not fully define $\tau$.  In either case, a model, which we call an
embedded model, is required, and is written $\tau_m$:
\begin{equation}
\label{eqn:embedded}
\tau \approx \tau_m(u; s, \theta),
\end{equation}
where $\approx$ indicates that the model is imperfect; $s$ is a set of
scenario variables for the embedded model; and $\theta$ is a set of
parameters required by the model in addition to the scenario. These are
the calibration or tuning parameters for the model. Note that the scenario space of the embedded
model may be different from that of the global model in which it is
embedded, so that $r$ and $s$ are not necessarily the same.  In
particular, $s$ often includes only a subset of the variables of $r$.
Further, in many settings, $\tau_m$ is formulated entirely in terms of
the local solution $u$ and calibration parameters $\theta$, in which
case $s$ is empty.
The fact that the scenario spaces of the global
composite model and the embedded model are different is an important
feature enabling reliable extrapolative predictions.

For the purposes of model calibration and validation, we require that
some observable quantities $y$ can be measured experimentally. These
observable quantities are different from the prediction QoIs $q$, but
both $y$ and $q$ are determined from the model state, the scenario, and
possibly the embedded
model:
\begin{gather}
y=\mcal{Y}(u,\tau;r), \label{eqn:maptoobs} \\
q=\mcal{Q}(u,\tau;r), \label{eqn:maptoqoi}
\end{gather}
where, for simplicity of exposition, the theories underlying the operators
$\mcal{Y}$ and $\mcal{Q}$ are presumed to be as reliable as the models
embodied by $\mcal{R}$.  

\subsection{Background on Model Discrepancy}
\label{sec:back_discrepancy}
In general, the model $\tau_m$ is less reliable than the model in
which it is embedded either because its dependencies or functional
form are incorrect (structural inadequacy/uncertainty) or because the
parameters $\theta$ are not perfectly known (parameter
error/uncertainty) or both.  These errors introduce error in the model
and, in turn, in both the solution and the QoI.  In their seminal
paper, Kennedy and O'Hagan~\cite{Kennedy2001} address this fact by
introducing a statistical model for ``the difference between the true
value of the real world process and the code output at the true values
of the inputs''.  In their approach, which has become very common in
the Bayesian literature~\cite{Higdon2005, Bayarri2007}, this
statistical model is posed directly for the observable quantities,
such that~\eqref{eqn:maptoobs} is replaced by
\begin{equation*}
y=\mcal{Y}(u,\tau_m;r) + \delta_y(r;\alpha),
\end{equation*}
where $\delta_y$ is the model for the uncertainty in the observables
induced by structural inadequacy, and $\alpha$ are a set of
hyperparameters for that model, while $u$ is still determined by the
original composite model $\mcal{R}(u, \tau_m; r) = 0$.  

In the present context, this treatment of the error is incomplete
because it says nothing about the effect of model discrepancy on the
QoIs.  An analogous formulation for the QoIs is
\begin{equation*}
q=\mcal{Q}(u,\tau_m;r) + \delta_q(r;\beta),
\end{equation*}
and thus, the full model is given by
\begin{gather}
\label{eqn:gpsystem}
\mcal{R}(u,\tau_m;r)=0, \\
y=\mcal{Y}(u,\tau_m;r) + \delta_y(r;\alpha), \\
q=\mcal{Q}(u,\tau_m;r) + \delta_q(r;\beta).
\end{gather}
Here, the hyperparameters $\alpha$ and $\beta$ need to be calibrated,
as well as the parameters of $\tau_m$, $\theta$. In the Kennedy and
O'Hagan approach, the calibration of $\theta$ and $\alpha$ would be
performed together using data for the observable $y$. But, because
data for the QoIs $q$ are not available, this approach cannot be used
to calibrate the hyperparameters $\beta$. Furthermore, even if one
were able to pose a model $\delta_q$ that did not require calibration,
there would be no way to test this model in the validation process,
making it inappropriate for use in predictions. Clearly, a different
approach is needed.  Here we propose to take advantage of the
structure of the composite model to introduce model uncertainty
representations that can be informed and tested using data as well as
used to quantify the uncertainty in predictions of the QoIs.

\subsection{Discrepancy Modeling for Predictive Validation} 
\label{sec:modeling}
The point of predictive validation is to assess the predictive
capability of the model---i.e., to characterize the accuracy of the
predictions of $q$.  A key challenge in this process is to determine
what the observed discrepancies between the model outputs and
observational data imply about the reliability of the QoI
predictions.  In the formulation of~\S\ref{sec:back_discrepancy}, one
must infer $\delta_q$ from observations of $y$.  In general, this
inference is extremely challenging because there is no direct mapping
from the observables to the QoIs---i.e., given only $y$, one cannot
evaluate $q$.  Thus $\delta_q$ cannot be constructed directly from the
physical model and the observations alone.  Additional modeling
assumptions are required.

Alternatively, in the predictive validation process, a mathematical
relationship between the observables and the QoIs is constructed by
formulating uncertainty models to represent errors at their sources.
Such models are able to provide uncertain predictions for both the
observables and the QoIs without additional assumptions.  Thus,
observational data can be used to inform and test these uncertainty
models, and that information can directly influence the predictions of
the QoIs.

To accomplish this, we recognize that the sole source of the modeling
error in the current example is the embedded model $\tau_m$.  Thus,
instead of modeling the effects of this error on the observables and
QoIs separately as in~\S\ref{sec:back_discrepancy}, we enrich the
embedded physical model~\eqref{eqn:embedded} with a model that
represents not only the physics of the original model but also the
uncertainty introduced by the structural inadequacy of that model.
For example, one could write
\begin{equation*}
\tau \approx \tau_m(u, s; \theta) + \epsilon_m(u, s; \alpha),
\end{equation*}
where $\epsilon_m$ denotes the uncertainty representation, which may
depend on additional parameters $\alpha$.  Given our choice to use
probability to represent uncertainty, it is natural that $\epsilon_m$
is a stochastic model, even when the physical phenomenon being modeled
is inherently deterministic.  Of course, an additive model is not
necessary; other choices are possible.   More importantly, the form of
$\epsilon_m$ must be determined.  The specification of a stochastic
model $\epsilon_m$ is driven by physical knowledge about the
nature of error as well as practical considerations necessary to make
computations with the model tractable.  
Although general principles for developing physics-based uncertainty
models need to be developed, the specification of such a model is
clearly problem-dependent and, thus, will not be discussed further
here.

For the current purposes, it is sufficient to observe that the model
$\epsilon_m$ is posed at the source of the structural
inadequacy---i.e., in the embedded model for $\tau$.  The combination
of the physical and uncertainty models forms an enriched composite model, which takes
the following form in the current case:
\begin{subequations}
\begin{gather}
\mcal{R}(u,\tau_m+\epsilon_m;r)=0, \\
y=\mcal{Y}(u,\tau_m+\epsilon_m;r), \\
q=\mcal{Q}(u,\tau_m+\epsilon_m;r).
\end{gather}
\label{eqn:uqsystem}
\end{subequations}
The inadequacy model, $\epsilon_m$, appears naturally in the calculation
of both $y$ and $q$, both directly through the possible dependence of
$\mcal{Y}$ and $\mcal{Q}$ on $\tau$, and indirectly via the dependence of
$u$ on $\tau$ through $\mcal{R}$. The structural uncertainty can
therefore be propagated to both the observables and the QoIs.
Furthermore, one can learn about $\epsilon_m$---i.e., inform and test
the model---from data on the observables and then transfer that
knowledge to the prediction of the QoIs. The predictive validation
process described in \S\ref{sec:predval} is designed to address prediction
problems of this type. The process involves training (calibration)
embedded models and their inadequacy models, testing (validation) the
embedded model and its use in the enriched composite model~\eqref{eqn:uqsystem},
and assessing (predictive assessment) whether the knowledge gained
through these processes can be reliably transferred to the QoI
predictions.

\subsection{Generalized Problem Abstraction for Complex
 Systems}\label{sec:genproblem}

The abstract problem statement from \S\ref{sec:problem} was designed
as a simple illustrative example of a broad class of problems in which
validated predictions of unobserved quantities are needed. This simple
formulation can be generalized in a number of ways to encompass most
computational prediction problems in science and engineering,
including predictions in complex physical systems, in which many
interacting physical phenomena are at work. These generalizations are
discussed here.

The first generalization is that many unreliable embedded models of
different phenomena may be required to represent the system being
studied. For example, in a combustion problem, models for
thermodynamics, molecular transport, radiation transport, and chemical
reactions may need to be embedded in the equations for the conservation
of mass, momentum and energy. To represent this situation, we introduce a set of
$N$ quantities $\tau_i$ for $1\le i\le N$ which need to be modeled to
close the governing equations:
\begin{equation}
\mcal{R}(u,\tau_1,\tau_2,\ldots,\tau_N;r)=0.
\end{equation} 
An embedded model for each of these quantities would then be needed,
each of which could have a set of calibration parameters $\theta_i$, a
set of model-specific scenario parameters $s_i$, and an inadequacy
representation $\epsilon_{im}$ with hyperparameters $\alpha_i$:
\begin{equation}
\tau_i\approx\tau_{im}(u;\theta_i,s_i)+\epsilon_{im}(u;\alpha_i,s_i).
\end{equation}

The second generalization is that the experimental situation in which
observations are made may be so different from the prediction
situation that it is represented by a different reliable model $\cal
R$. This experimental scenario might also depend on a number of new
quantities $\tau$ that must be modeled. For example, these might
represent the response of the measuring instrument or some
characteristic of the experimental apparatus. In this case, the
experimental scenario may be described by different parameters than
the prediction scenario. There will in general be some number $N_e$ of
different experiments, each of which involves a set of modeled
quantities particular to the experiment as well as at least one of the
modeled quantities used in predictions.  To provide the most direct
information regarding the embedded models used in the predictions,
each of the experimental scenarios used for calibration would involve
only one of the embedded models used for prediction and a minimum
number of additional embedded models required to simulate the
experiment (ideally none).  

This situation can be formalized by introducing $N_e$ theoretical
descriptions $\mathcal{R}^i$, where the superscript is the experiment
index. Each of these models depends on a set $\{\tau\}^i$ of the
embedded models used for prediction, with elements $\tau^i_j=\tau_k$
for some $k$, and a set of additional embedded models necessary for
the experiment, denoted by $\{\sigma\}^i$ with elements
$\sigma^i_j$. Furthermore, each of the experiments will in general
involve a different observation model $\mathcal{Y}^i$. The experiments
to be used for calibration and validation can therefore be expressed:
\begin{eqnarray}
0 &=& \mathcal{R}^i(u^i,\{\tau\}^i,\{\sigma\}^i;r^i) \qquad 1\le i\le N_e\\
y^i &=& \mathcal{Y}^i(u^i,\{\tau\}^i,\{\sigma\}^i;r^i) \qquad 1\le
 i\le N_e
\end{eqnarray}
where $u^i$ denotes the state variables, $y^i$ is the vectors of
observation data, and $r^i$ is the vector of scenario parameters for
experiment $i$.

Each of the prediction modeled quantities $\tau_k$ has been modeled as
$\tau_{km}+\epsilon_{km}$ for use in the prediction, as discussed above, so
each vector of prediction modeled quantities $\{\tau\}^i$ is associated with
a vector of models $\{\tau_m\}^i$ and a vector of error models
$\{\epsilon_m\}^i$. In addition, models for the experimental modeled
quantities $\sigma$ must be posed. For each experiment $i$, there is thus
a vector of models $\{\sigma_m\}^i$, and each such model may have an
associated error model, so there is generally a vector of error models
$\{\delta_m\}^i$ for each experiment. These models will in general have
parameters which must be determined from data, and their validity will
need to be assessed as with the prediction models. 

The final generalization arises because the state variables $u^i$
associated with model $\mathcal{R}^i$ for the various experiments need
not be the same, or consistent with the state variables $u$ for the
prediction model. For example, in a fluid dynamics problem, the
prediction state variables would generally be a three-dimensional vector
field, while in the model for the viscometer in which the parameter in
the constitutive model for the internal stress (the viscosity) is
determined, the state variable is a one-dimension function for the
azimuthal velocity. Thus while the same embedded model is applied in
both the prediction and the experiment, the dependence of the embedded
model on the state must be different. We can express this by defining
arguments $v_k$ for each embedded model that are consistent for the
prediction and all experiments. An operator $\mathcal{V}^i_k$ is needed
that maps the state variable for each scenario to the argument of the
model $\tau_{km}$.

With these generalizations, the abstract statement of the prediction
problem, analogous to~\eqref{eqn:uqsystem} is
\begin{eqnarray}
0 &=& \mathcal{R}(u,\{\tau_m\}^0+\{\epsilon_m\}^0,r)\\
q &=& \mathcal{Q}(u,\{\tau_m\}^0+\{\epsilon_m\}^0,r)\\
0 &=& \mathcal{R}^i(u^i,\{\tau_m\}^i+\{\epsilon_m\}^i,\{\sigma_m\}^i+\{\delta_m\}^i,r^i)\qquad \mbox{for $1\le
 i\le N_e$}\\
y^i &=& \mathcal{Y}^i(u^i,\{\tau_m\}^i+\{\epsilon_m\}^i,\{\sigma_m\}^i+\{\delta_m\}^i,r^i)\qquad \mbox{for $1\le
 i\le N_e$}
\end{eqnarray}
where the embedded models appearing in composite model of the prediction
have the form:
\begin{equation}
\tau_k\approx\tau_{km}(\mathcal{V}_k(u),\theta_k,s_k) +
 \epsilon_{km}(\mathcal{V}_k(u),\alpha_k,s_k)
\end{equation}
while when the same models appear in the composite model for experiment
$i$ they have the form:
\begin{equation}
\tau_k\approx\tau_{km}(\mathcal{V}^i_k(u),\theta_k,s_k) +
 \epsilon_{km}(\mathcal{V}^i_k(u),\alpha_k,s_k).
\end{equation}

Clearly, if the experiments are to provide any meaningful information
about the prediction models, then the errors and uncertainties
associated with the embedded models for $\{\sigma\}^i$ will have to be
sufficiently small, or ideally entirely absent. That is, it is
preferable if there are no extra modeled quantities in the composite
model for the experiment. Similarly, as mentioned above, experiments in
which only one of the embedded models is exercised are particularly
valuable for learning about that model, since it avoids confounding
uncertainties from other models. Experiments that exercise many embedded
models are generally not as useful for calibrating embedded models.

This fact leads to the idea of a validation
pyramid\cite{Babuska_2007}.  In particular, it is helpful to organize
the experimental inputs to the simulation of a complex system
hierarchically. At the lowest level of the hierarchy are simple
experiments that exercise only one, or few embedded models. Such
experiments are generally relatively inexpensive, relatively well
controlled and numerous. They therefore can provide abundant well
characterized data, which is ideal for the calibration of embedded
models. These experiments are at the base of the pyramid. As one
ascends the hierarchy, or pyramid, the experiments exercise more of
the embedded models; they become increasingly expensive, and they
commonly become more difficult to control and instrument. Data from
these more complex experiments are thus more limited and are often of
lower quality, that is, they have higher uncertainty. For these
reasons, the higher experiments are in the hierarchy, the less useful
they are for calibrating embedded models. They are, however, critical
for validation testing. At the highest level of the pyramid are
experiments conducted on systems with complexity similar or
identical to the system of interest. Experiments on the system of
interest are particularly valuable because they provide an opportunity
to detect unanticipated and therefore unmodeled phenomena.

The generalizations discussed above are important to understanding how
the predictive validation process discussed here applies to the
simulation of complex systems. However, the basic ideas underlying
predictive validation are more easily discussed in the context of
simpler examples, which are well characterized by the simple abstract
problem described in \S\ref{sec:problem}. Therefore, the
generalizations described above will not be discussed further.

\section{Predictive Validation Processes}
\label{sec:predval}

Given a composite model like~\eqref{eqn:uqsystem}, there are likely to
be parameters---e.g., $\theta$ and $\alpha$ in $\tau_m$ and
$\epsilon_m$, respectively, from~\S\ref{sec:modeling}---that must be
determined from observations (calibration). If the inadequacy model
$\epsilon_m$ faithfully represents the discrepancies between the model
for the observables $\mcal{Y}$ and the observations (validation), we
then use it in the model for the QoIs $\mcal{Q}$ to predict how the
observed discrepancies impact uncertainty in the QoI, and assess the
adequacy of the calibration and validation processes for the QoI
prediction (predictive assessment).  Thus, the process needed to assess
the accuracy and credibility of the predictions involves three
activities: calibration, validation, and predictive assessment.  These
activities are described briefly below.

\subsection{Model Calibration}
The parameters in the embedded models $\tau_m$ and inadequacy models
$\epsilon_m$ need to be specified in some way. Some parameters may be
very well known, with either known or negligible uncertainties, and
such parameters need not be calibrated (e.g., the speed of light or
the acceleration due to gravity on Earth). However, in most cases, at
least some of the parameters are not well-known, and so values must be
determined that are consistent with existing knowledge about the
phenomenon and with observational data. Generally, this can be posed
as an inverse problem, where the model inputs are be determined by
requiring the from the model outputs to match observations under
constraints imposed by prior knowledge. Furthermore, the uncertainties
in the data being used and the qualitative and often imprecise nature
of existing knowledge about the parameters must be accounted for to
yield estimates of the resulting uncertainty in the calibrated
parameters.

There are a number of different approaches that might be formulated to
solve such inverse problems \cite{Biegler2010}.
Provided they produce consistent estimates and uncertainties
for the parameters, they can serve as calibration techniques for the
predictive validation approach discussed here.
Recall, however, that we have selected Bayesian probability for our
mathematical representation of uncertainty. In this context, a very natural
and powerful approach to calibration is Bayesian inference, which
relies on Bayes' theorem:
\begin{equation}
p(\theta, \alpha | Y, I) = \frac{ p(\theta, \alpha | I) L(\theta,
 \alpha; Y, I) }{\int p(\theta, \alpha|I) L(\theta, \alpha; Y,I) \, d\theta d\alpha}
\end{equation}
where $Y$ represents the calibration data, and $p(\cdot|\cdot)$ is a
conditional probability density function (PDF). 
Here, $p(\theta, \alpha | I)$ is the joint prior PDF of the parameters
$\theta$ and $\alpha$, which is conditional on the prior knowledge $I$.
The function $L(\theta, \alpha; Y,I)$ is the
likelihood, which is given by:
\begin{equation}
L(\theta, \alpha; Y, I) = p(y=Y|\theta, \alpha, I).
\label{eq:likelihood}
\end{equation}
It represents the likelihood of observing the data given the model and
its uncertainty, the particular values of $\theta$ and $\alpha$, and
the uncertainty in the measurements.
Finally, $p(\theta, \alpha | Y, I)$ is the posterior PDF, the joint distribution
of the parameters, conditioned on both the data and the prior
knowledge. This posterior PDF is the Bayesian solution of the inverse problem, representing
the desired estimate of the parameters, with uncertainties.

\subsection{Validation} \label{sec:methodology_validation}
Calibrating the model does not guarantee that its outputs will be
consistent with the calibration data, much less new data not used for
calibration. Indeed, calibration can only ensure that the model matches the
calibration data as well as possible, which may not be very well at
all. It is therefore necessary that consistency of model outputs with
the experimental observations be explicitly checked. This process of
comparing models to observations is consistent with the classical notion
of validation. However, in making such comparisons, determining how much
discrepancy between model and observations is acceptable is subtle and
important. The most relevant metric is the implied discrepancy in
the QoI (i.e., the difference between the predicted QoI and reality),
but this metric is inaccessible in standard validation approaches.
Hence, the determination of a tolerance is generally left to expert
opinion \cite{Oberkampf_2010}.

However, the consideration of uncertainties provides a natural way to
define the acceptable level of discrepancy. It is simply that the data
must be a plausible outcome of the model, with all its uncertainties,
including those due to the model inputs, observational errors, and model
inadequacy. This simply tests whether the uncertainty models,
particularly the inadequacy model, can plausibly account for the causes
of the discrepancies. If any of the available data, including data used
for calibration as well as that intended only for validation, is not a
plausible outcome of the model with its uncertainties, then we can
conclude that the uncertainty representations are somehow insufficient
and should not be used for prediction.

It remains then to define what it means for the data to be a plausible
outcome of the model. This clearly depends on how uncertainties are
being represented mathematically. If one represents uncertainty using
probability, as we do here, then the validation criterion will clearly
need to be that the data is not too improbable.  To quantify this we
require a metric and a tolerance.

Given the probability distributions for the model inputs $\theta$ and
$\alpha$ obtained from the calibration process, the
model \eqref{eqn:uqsystem} yields a probability density for each
individual observable quantity $y_i$, $p(y_i|Y,I)$. This probability
density is formally conditional on all the calibration data and our
prior knowledge, due to the calibration process, and it is helpful to
explicitly acknowledge this fact, as is done here. Let an
observational datum for observable $y_i$ be $Y_i$, then a useful
approach to determine whether the datum is a plausible outcome of the
model can be based on Bayesian credible intervals, of which several
can be defined.  

Particularly appropriate for our use are highest
posterior density (HPD) credibility intervals \cite{Box1973}.  The
$\beta$-HPD $(0 \leq \beta \leq 1)$ interval is an interval or a set of intervals 
for which the probability of belonging to it is $\beta$ and 
the probability density of all the points in the interval is greater than the density of 
the points outside the interval. However, because HPD credibility
sets are defined in terms of the probability density, they are not
invariant to a change of variables. This is particularly
undesirable when formulating a validation metric because it means that one's
conclusions about model validity would depend on the arbitrary
choice of variables (e.g., whether one considers the observable to be
the frequency or the period of an oscilation). To avoid this problem, we
introduce a modification of the HPD in which the credible set is defined
in terms of the probability density relative to a specified distribution
$q$. An appropriate definition of $q$ would be the ignorance
distribution~\cite{Jaynes_2003} on the data space.
Using this definition of the highest posterior \emph{relative}
density (HPRD), a credibility metric can be defined as
$\gamma=1-\beta_{\rm min}$, where $\beta_{\rm min}$ is the smallest value
of $\beta$, for which the observation $Y_i$ is in the HPRD-credibility
set for $y_i$. That is:
\begin{equation}
\gamma(Y_i)=1-\int_S p(y_i|Y,X)\,dy_i, \quad\mbox{where}\quad 
S=\left\{y_i:\frac{p(y_i|Y,X)}{q(y_i)}\geq
   \frac{p(Y_i|Y,X)}{q(Y_i)}\right\}
\label{eq:gammaHPRD}
\end{equation}
When $\gamma$
is smaller than some tolerance, say less than $0.05$ or $0.01$, the
data are considered an implausible outcome of the model.

%
%
A validation metric based on HPRD credibility sets is proposed
here because it
seems to be best aligned with intuitive notions of agreement between the
model predictions and observations, even for skewed and multi-modal
distributions.  For example, for multi-modal distributions, an HPRD
region may consist of multiple disjoint intervals~\cite{Hyndman1996}
around the peaks in the distribution, leaving out the low probability density regions
between the peaks.  

Even if all available data are credible outcomes of the model by
an HPRD criterion, this
does not imply that the model with its associated uncertainty is valid
for predicting the QoIs. Indeed it only implies that we have so far
failed to invalidate the model for use in making predictions. To
determine the validity of a prediction, it is necessary to assess
whether the validation tests have been sufficiently thorough to give
confidence in the prediction, and whether, in the prediction scenario,
all the embedded models are being used in regimes in which they are
expected to be reliable. This is the role of predictive assessment.

\subsection{Predictive Assessment}
\label{sec:predict_assess}

As discussed in \S\ref{sec:introduction}, assessing the reliability of
a prediction is difficult because prediction fundamentally requires
extrapolation from available data. Such an extrapolation cannot be
justified based on the agreement between the data and the model
alone. Knowledge about the structure of the physical and uncertainty
models and their potential shortcomings is important to determining
whether the validation tests that have been performed are sufficient
to warrant confidence in the predictions. There are two primary
questions that need to be addressed. The first is whether the QoIs are
sensitive to aspects of the embedded models that have not been
effectively informed by the calibration and tested through the
validation process.  The second is whether the embedded models and
their accompanying inadequacy models are used 
outside their ``domain of applicability.''  Answering these
questions is central to assessing the credibility of the prediction
process.

Because these credibility assessments rely so heavily on knowledge of
the characteristics and pedigree of the embedded models, it is not
possible to provide general prescriptions for performing these
assessments. Instead, in the following subsections, a number of important
considerations will be discussed in the context of examples.

In addition, if a prediction is determined to be credible, there is
finally the question of whether the prediction is sufficiently precise;
that is, whether it has sufficiently small uncertainty for the purposes
for which it is being performed. While this is important, it clearly depends
on the nature of the decisions to be informed by the predictions. It is
therefore out of scope for the current discussion, and will not be
discussed further here.

\subsubsection{Adequacy of Calibration \& Validation}

The fundamental issue in assessing the adequacy of the calibration and
validation, is whether QoIs are sensitive to some characteristic of an
embedded model while the available data are not. The prediction then
depends in some important way on things that have not been informed or
constrained by the data. In this case, the prediction can only be
credible if there is other reliable information that informs this aspect
of the embedded models. To assess this then requires a sensitivity
analysis to identify what is important about the embedded models for
making the predictions. To make this generic discussion more concrete,
it is useful to consider several representative cases.

\paragraph{Sensitivity to a Model Parameter} Suppose that the prediction
QoI is highly sensitive to the value of a particular parameter
$\theta$ in an embedded model. In this case, it is important to
determine whether the value of this parameter is well constrained by
reliable information. If, for example, none of the calibration data
has informed the value of $\theta$, then only other available
information (prior information in the Bayesian context) has determined
its value. Further, if none of the validation quantities are sensitive
to the value of $\theta$, then the validation process has not tested
whether the information used to determine $\theta$ is in fact valid in
the current context. The prediction QoI is then being determined to a
significant extent by the untested prior information used to determine
$\theta$.  This should leave us little confidence in the prediction,
unless the prior information is itself highly reliable (e.g., $\theta$
is the speed of light). Alternatively, when the available prior information
is questionable (e.g., $\theta$ is the reaction rate of a poorly
understood chemical reaction), the predictions based on $\theta$ will
not be reliable.

Alternatively, the calibration data could have been highly informative
of the value of $\theta$ during the calibration process and the
validation quantities could have also been very sensitive to the value
of $\theta$. This would suggest that the value of the QoI is being
substantially determined by the calibration and validation data through
the sensitive parameter $\theta$. Provided the data are reliable, the
determination of $\theta$ will not cause the prediction to be unreliable.

\paragraph{Sensitivity to an Embedded Model}
Suppose that the prediction QoI is highly sensitive to one of the
embedded models $\tau_m$. In this case, it is important to determine
whether the embedded model and its use in the composite model has been
well validated. If, for example, none of the validation quantities are
sensitive to $\tau_m$, then the validation process has not provided a
test of the validity of $\tau_m$, and a prediction based on $\tau_m$
would be unreliable. However, such a clear situation is not likely,
especially when $\tau_m$ involves parameters that have been
calibrated, usually with scenarios low on the validation pyramid
(see \S\ref{sec:genproblem}), because at least the calibration data
would then be expected to be sensitive to $\tau_m$. A more common
situation would be that validation quantities from scenarios higher on
the pyramid are not sensitive to $\tau_m$, so that the validity of
using $\tau_m$ in a composite model similar to that used in the
predictions has not been tested.

To see why this could be problematic, consider an embedded model
$\tau_m$ for a quantity $\tau$, which in the prediction scenario depends
on one of the system state variables $u$, and suppose that this
dependency has not been recognized so that $u$ is not included as an
argument in $\tau_m$. In
the simplified scenarios near the bottom of the validation pyramid that
are used to calibrate and validate $\tau_m$, this dependence might be
absent or unimportant. So this data could not be used to detect the
missing dependence. If the validation quantities in tests that are
higher on the pyramid, in which the omitted dependence is important, are
insensitive to $\tau_m$, then these test will also not detect the
omission. To guard against this and similar possible failures of
$\tau_m$, the predictive assessment process should determine whether
validation quantities in scenarios ``close enough'' to the prediction
scenario are sufficiently sensitive to $\tau_m$ to provide a good test
of its use in the prediction. The determination of what is ``close
enough'' and what constitutes sufficient sensitivity must be made based
on knowledge of the model and the approximations that went into it, and
of the way the models are embedded into the composite models of the
validation and prediction scenarios. If there are no data for
sufficiently sensitive quantities on close enough scenarios, then the
resulting predictions would be unreliable, unless one had independent
information that the model is trustworthy for the prediction.

\paragraph{Sensitivity to an Inadequacy Model}
Suppose that uncertainty in the prediction QoI is largely due to the
uncertainty model $\epsilon_m$ representing the inadequacy of the
embedded model $\tau_m$. In this case, it is important to ensure that
$\epsilon_m$ is a valid description of the inadequacy of $\tau_m$. As
with the embedded model sensitivities discussed above, validation tests
from high on the validation pyramid are most valuable for assessing
whether the uncertainty model represents inadequacy in the context of a
composite model similar to that for the prediction. If however, the
available validation data are for quantities that are insensitive to
$\epsilon_m$, then the veracity of $\epsilon_m$ in representing the
uncertainty in the QoI will be suspect. Reliable predictions will then
be possible only if there is independent information that the
inadequacy representation is trustworthy.

\subsubsection{Domain of Applicability of Embedded Models}

In general, it is expected that the embedded models making up the
composite model to be used in a prediction will involve various
approximations and/or will have been informed by a limited set of
calibration data. This will limit the range of scenarios for which the
model can be considered reliable, either because the approximations
will become invalid or because the model will be used outside the
range for which it was calibrated. It is therefore clearly necessary
to ensure that the embedded models are being used in a scenario regime
in which they are expected to be reliable.

As discussed in \S\ref{sec:problem}, reliable extrapolative
predictions are possible because the scenario parameters relevant to
an embedded model need not be the same as those for the global composite
model in which it is embedded. For example, when modeling the
structural response of a building, the scenario parameters include the
structural configuration and the loads. However, the scenario
parameters for the linear elasticity embedded model used for the internal
stresses would be the local magnitude of the strain, as well as other
local variables such as the temperature. For each embedded model then,
we need to identify the scenario parameters that characterize the
applicability of the model and the range of those parameters over which
the model and its calibration is expected to be reliable. It is then a
simple matter of checking the solution of the composite model to see
if any of the embedded models are being used ``out of range''. For some
embedded models, defining the range of applicability in this way is
straightforward, but for others this represents a challenging
research issue.

\subsubsection{Other Issues}

There are two other issues that must be considered when performing a
predictive assessment and interpreting the results of a
prediction. First, the focus of the discussion to now has been on
ensuring that the calibration and validation processes have been
sufficiently rigorous to warrant confidence in an extrapolative
prediction and its uncertainty. However, a prediction with an
uncertainty that is too large to inform the decision for which the
prediction is being performed is not sufficient, even if that
uncertainty has been determined to be a good representation of what
can be predicted about the QoI. The requirements for prediction
uncertainty to inform a decision based on the prediction depend on the
nature of the decision, and determination of this requirement is
outside the scope of the current discussion. However, once such a
requirement is known, the prediction uncertainties can be checked to
determine whether these requirements are met, and therefore whether the
prediction is useful.

The second issue is the well-known problem of ``unknown unknowns.'' If
the system being simulated involves an unrecognized phenomenon, then
clearly an embedded model to represent it will not be included in the
composite model for the system. As with the examples above, the
prediction QoI could be particularly sensitive to this phenomenon,
while the validation observables are not sensitive. In this situation,
one would not be able to detect that anything is missing from the
composite model.  Further one could not even identify that the
validation observables were insufficient; that is, the predictive
assessment could not detect the inadequacy of the validation
process. The is a special case of a broader issue. The predictive
validation process developed here relies explicitly on reliable
knowledge about the system and the models used to represent it. This
knowledge is considered to not need independent validation, and is
thus what allows for extrapolative predictions. However, if this
externally supplied information is in fact incorrect, then the
predictive validation process may not be able to detect it.

\section{Illustrative Example} 
\label{sec:results}
To illustrate some aspects of the predictive validation process,
we apply it here to a simple problem involving a spring-mass-damper system.
For this system, Newton's second law requires that
\begin{equation}
m \ddot{x} = f_d + f_s,
\label{eqn:fma}
\end{equation}
where $m$ is the mass, $x$ is the position of the mass, and $f_d$ and
$f_s$ are the forces acting on the mass due to the damper and the
spring, respectively.  For this example, other forces acting on the
mass (e.g., drag, gravity) are known to have negligible effect.  Thus,
\eqref{eqn:fma} represents highly reliable theory in the context of
the current problem.  However, $f_d$ and $f_s$ must be
specified---i.e., potentially inadequate embedded models are required
for these forces.  For the purposes of this example, a truth system,
detailed in Appendix~\ref{app:truth}, is used to make simulated
experimental observations but is otherwise unknown to the modeler.
The goal of the predictive validation exercise is to evaluate the use
of simple linear models for $f_d$ and $f_s$ to predict the QoI, which
is taken to be the maximum velocity of the mass for $m=5$ with an
initial condition of $x(0) = 4$, $\dot{x}(0) = 0$.

Execution of the predictive validation process depends on our ``state
of knowledge''---that is, the available information about the system.
The information used here is described in \S\ref{sec:information}.  In
the current exercise, three different states of knowledge are
considered, which lead to different results.  In particular, the
different states of information lead to different uncertainty
representations, which are discussed in~\S\ref{sec:uncertainty_reps},
as well as different validation procedures and conclusions, which are
described in~\S\ref{sec:numerical_results}.

\subsection{Available Information} \label{sec:information}
In general, the information available in the prediction process is of
several types.  In the present problem, the available information
consists of the high fidelity model~\eqref{eqn:fma}, the composite
physical model which will be specified below, information regarding the
inadequacies of the embedded models for the forces, and observational
data.

\subsubsection{The Composite Model}
The standard mathematical representations for both springs and dampers
are linear, and since no information is available to suggest a
specific nonlinear models, linear models will be used here. This is
clearly a modeling assumption, and must be considered provisional.
The embedded models are thus
\begin{align*}
f_s &= -k x, \\
f_d &= -c \dot{x},
\end{align*}
where the constants $k$ and $c$
are unknown. With these embedded models, the composite model is
\begin{gather}
m \ddot{x} + c \dot{x} + kx = 0, \label{eqn:composite}\\
x(0) = 4, \quad \dot{x}(0) = 0. \nonumber
\end{gather}
Note that the composite model~\eqref{eqn:composite} is not the same as
the truth model described in Appendix~\ref{app:truth}.  Thus, it is
necessary to consider model inadequacy in the predictive validation
exercise.  Further, information about the model inadequacy that is
known independently of the observational data that will be used for
calibration and validation will be necessary.

To demonstrate how predictive validation depends on the available
information, we postulate three different states of knowledge
regarding the inadequacy of the model.  These are labeled States of
Information 0, 1, and 2 or SI0, SI1, and SI2 for short.  In SI0, we
are confident that physical model \eqref{eqn:composite} is sufficient
for the required prediction---i.e., that the embedded models are
highly accurate and that there are no important neglected physics.
However, the parameter values $k$ and $c$ are not well-known and must
be calibrated from the observational data.

In SI1, it is known that all important forces are represented and that
the linear spring model is highly accurate, while the the constant
coefficient damping model is suspected to be inadequate.  However, no
information is available about why the damping model might fail to represent
reality.  

Finally, in SI2 (as in SI1), the linear spring model is known to be
adequate, and the damping model is known to be problematic.  Unlike SI1,
we have information regarding the cause of the inadequacy.  It was
noticed that the damper becomes warm when it moves, presumably because
of the energy that is being dissipated by the damper. Because the
viscosity of the damping fluid in the damper can reasonably be assumed to
depend on temperature, this would result in a change in the damping
coefficient with time.  While the precise form of the model
inadequacy---i.e., the temperature variation of the damping fluid or the
temperature dependence of the damping coefficient---is not known, this
additional information will be essential in building confidence in
extrapolative predictions.

\subsubsection{Observational Data}
\label{sec:data}
To calibrate and validate the composite model, observations of the
position of the mass at discrete times for two different masses, $m=1$
and $m=2$, are available.  These observations are given in
Tables~\ref{tbl:data_m1} and~\ref{tbl:data_m2}.
\begin{table}[ht]
\caption{Observations of the actual system for $m=1$.}
\begin{center}
\begin{tabular}{|c|c|c|}
\hline
Time $(t)$ & Position $(x)$ & Observation $(x_{\mathrm{obs}} = x + \epsilon)$\\
\hline
0.0 & 4.0 & 4.0 \\
1.0 & $4.025647 \times 10^{-1}$ & $4.056287 \times 10^{-1}$ \\
2.0 & $-1.913556 \times 10^0$ & $-1.917800 \times 10^{0}$ \\
3.0 & $7.536144 \times 10^{-2}$ & $7.331597 \times 10^{-2}$ \\
4.0 & $8.219699 \times 10^{-1}$ & $8.176825 \times 10^{-1}$ \\
5.0 & $-1.260000 \times 10^{-1}$ & $-1.129453 \times 10^{-1}$ \\
6.0 & $-3.076154 \times 10^{-1}$ & $-3.011407 \times 10^{-1}$ \\
7.0 & $8.109402 \times 10^{-2}$ & $9.303637 \times 10^{-2}$ \\
8.0 & $1.062484 \times 10^{-1}$ & $8.884368 \times 10^{-2}$ \\
\hline
\end{tabular}
\end{center}
\label{tbl:data_m1}
\end{table}
\begin{table}
\caption{Observations of the actual system for $m=2$.}
\begin{center}
\begin{tabular}{|c|c|c|}
\hline
Time $(t)$ & Position $(x)$ & Observation $(x_{\mathrm{obs}} = x+\epsilon)$\\
\hline
0.0 & 4.0 & 4.0 \\
1.0 & 1.718579 & 1.705548 \\
2.0 & -1.641053 & -1.634634 \\
3.0 & -2.121425 & -2.127946 \\
4.0 & $-1.641898 \times 10^{-1}$ & $-1.818642 \times 10^{-1}$ \\
5.0 & 1.278992 & 1.269814 \\
6.0 & $8.442413 \times 10^{-1}$ & $8.507498 \times 10^{-1}$ \\
7.0 & $-3.168699 \times 10^{-1}$ & $-3.259694 \times 10^{-1}$ \\
8.0 & $-7.066765 \times 10^{-1}$ & $-7.080865 \times 10^{-1}$ \\
\hline
\end{tabular}
\end{center}
\label{tbl:data_m2}
\end{table}
The tables give both the actual position, as determined by solving the
truth system (see Appendix~\ref{app:truth}) using Runge-Kutta-Fehlberg
(4,5) time marching, and the data used here, which
are perturbed by simulated observation noise.  The observation noise is
such that the $i$th observed value $\hat{x}_i$ is given by
\begin{equation*}
\hat{x}_i = x_i + \epsilon_i,
\end{equation*}
where $x_i$ is the actual position and $\epsilon_i$ are independent,
identically distributed Gaussian random variables with mean zero and
standard deviation $0.01$.
 
\subsection{Uncertainty Representations for Calibration and Prediction}
\label{sec:uncertainty_reps}
The formulation of the uncertainty representation depends on the state of
knowledge.  Here we develop two different models: one corresponding to
SI0 and one corresponding to SI1 and SI2.

In SI0, we are confident that the physical model is adequate.  Thus,
no model inadequacy representation is required.  However, the values
of $k$ and $c$ are unknown.  Thus, following the standard Bayesian
approach, $k$ and $c$ are treated as random variables, which will be
calibrated using Bayesian inference, using the data given in
Table~\ref{tbl:data_m1}. For this, a joint prior probability of $k$
and $c$ must be specified, and for simplicity they are taken to be
independent with $k \sim N(1, 1/2)$ and $c \sim \log N(0, 1/2)$, where
$N(\mu, \sigma^2)$ denotes the Gaussian with mean $\mu$ and variance
$\sigma^2$, and $\log N(\mu, \sigma^2)$ denotes the corresponding
log-normal.

Since the composite model is assumed to be perfect except for the
parameter uncertainty, the likelihood function is based on the
experimental uncertainty alone.  Thus,
\begin{equation*}
L(k, c; \hat{x}_1, \ldots \hat{x}_N) = \prod_{i=1}^N \ell(k, c; \hat{x}_i),
\end{equation*}
where $\ell$ is the likelihood for a single data point.  Specifically,
\begin{equation}
\ell(k, c; \hat{x}_i) = 
\frac{1}{\sqrt{2 \pi \sigma^2}} 
\exp \left[ -\, \frac{1}{2} \frac{(\hat{x}_i - x_m(t_i; k,c))^2}{\sigma^2} \right],
\label{eqn:single_pt_likelihood}
\end{equation}
where $\sigma = 0.01$ is the standard deviation of the observation
error $\epsilon_i$ and $x_m(t_i; k,c)$ is the solution
of~\eqref{eqn:composite} at $t_i$ for parameters $k$ and $c$.

In both SI1 and SI2, the damping model is known to be inadequate. There
are many possible ways one might represent this inadequacy.  Here we
adopt a very simple model designed to represent the variability required
in the damping coefficient to reproduce the departure from constant
coefficient behavior.  
In particular, the damping coefficient is modeled as a random variable
with the randomness intended to describe the variability in the
damping needed to encompass the data, not only a lack of knowledge
about the true value of $c$.  Since the damper coefficient $c$ must be
positive, the variability of the damper modeled by a log-normal
distribution: $c \sim \log N(c_{\mu}, c_{\sigma}^2)$.  The parameters
$c_{\mu}$ and $c_{\sigma}$ are uncertain and thus must be learned from
the calibration data.  Note, however, that this calibration is
fundamentally different from that pursued for SI0.  For SI0, the
assumption is that there is a unique true value of $c$ that is to be
determined.  In this situation, the posterior PDF for $c$
characterizes uncertainty about this best value due to having only a
finite amount of uncertain data.  As the number of independent data
points increases, all of the probability will converge to a single
value of $c$.  Alternatively, for SI1 and SI2, the goal is to find
both the values of $c$ that are most likely and how much ``variation''
is necessary to cover the data.  Even in the limit of infinite data,
$c$ will be characterized by a log-normal distribution with non-zero
variance.

Of course, one could pose more complex models that might better describe
the actual inadequacy of the embedded damping model.  The development
of uncertainty representations is highly problem dependent, and
developing a more complex model here does not further the goal of
illustrating the process.  Thus, we consider only this simple inadequacy
model.

In the prior, the parameters $k$, $c_{\mu}$, and $c_{\sigma}$ are taken to be
independent, and the following marginal prior
distributions are chosen:
\begin{gather*}
p(k) = \log N(1, 0.5), \\
p(c_{\mu}) = N(0, 0.5), \\
p(c_{\sigma}) = \log N(-1, 0.4).
\end{gather*}

To form the likelihood, each data point is considered
an independent draw from the random model of the damping coefficient,
which is consistent with the ensemble model discussed above.  Thus, likelihood for
each data point is given by
\begin{equation*}
\hat{\ell}(k, c_{\mu}, c_{\sigma}; \hat{x}_i ) = 
\int_{0}^{\infty} \ell(k,c; \hat{x}_i) \, p(c | c_{\mu}, c_{\sigma}) \, dc,
\end{equation*}
where $p(c | c_{\mu}, c_{\sigma})$ is the log-normal PDF and
$\ell(k,c; \hat{x}_i)$ is as given
in~\eqref{eqn:single_pt_likelihood}.  Since the data points are
independent, the full likelihood is given by
\begin{equation*}
L(k, c_{\mu}, c_{\sigma}; \hat{x}_1, \ldots, \hat{x}_n) = 
\prod_{i=1}^{n} \hat{\ell}(k, c_{\mu}, c_{\sigma}; \hat{x}_i ).
\end{equation*}
%

\subsection{Numerical Results}
\label{sec:numerical_results}

Predictive validation is necessarily pursued in the context of our state
of information, so each different state
of information is considered separately below.  In each case, the
actions taken as part of the predictive validation process and
the results observed are described.

\subsubsection{State of Information 0}
\begin{figure}[htp]
\begin{center}
\subfloat[$k$]{\includegraphics[width=0.43\linewidth]{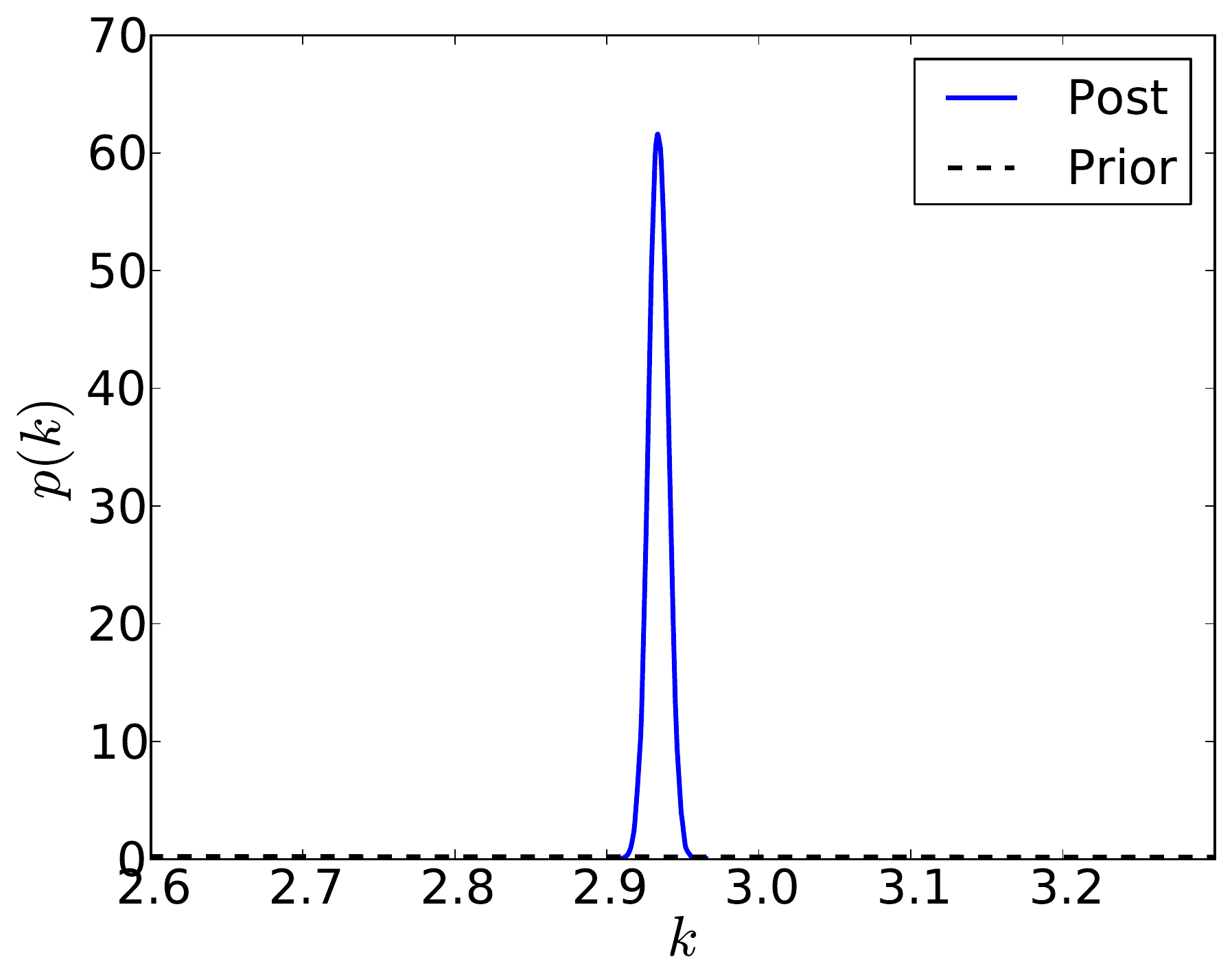}}
\subfloat[$c$]{\includegraphics[width=0.45\linewidth]{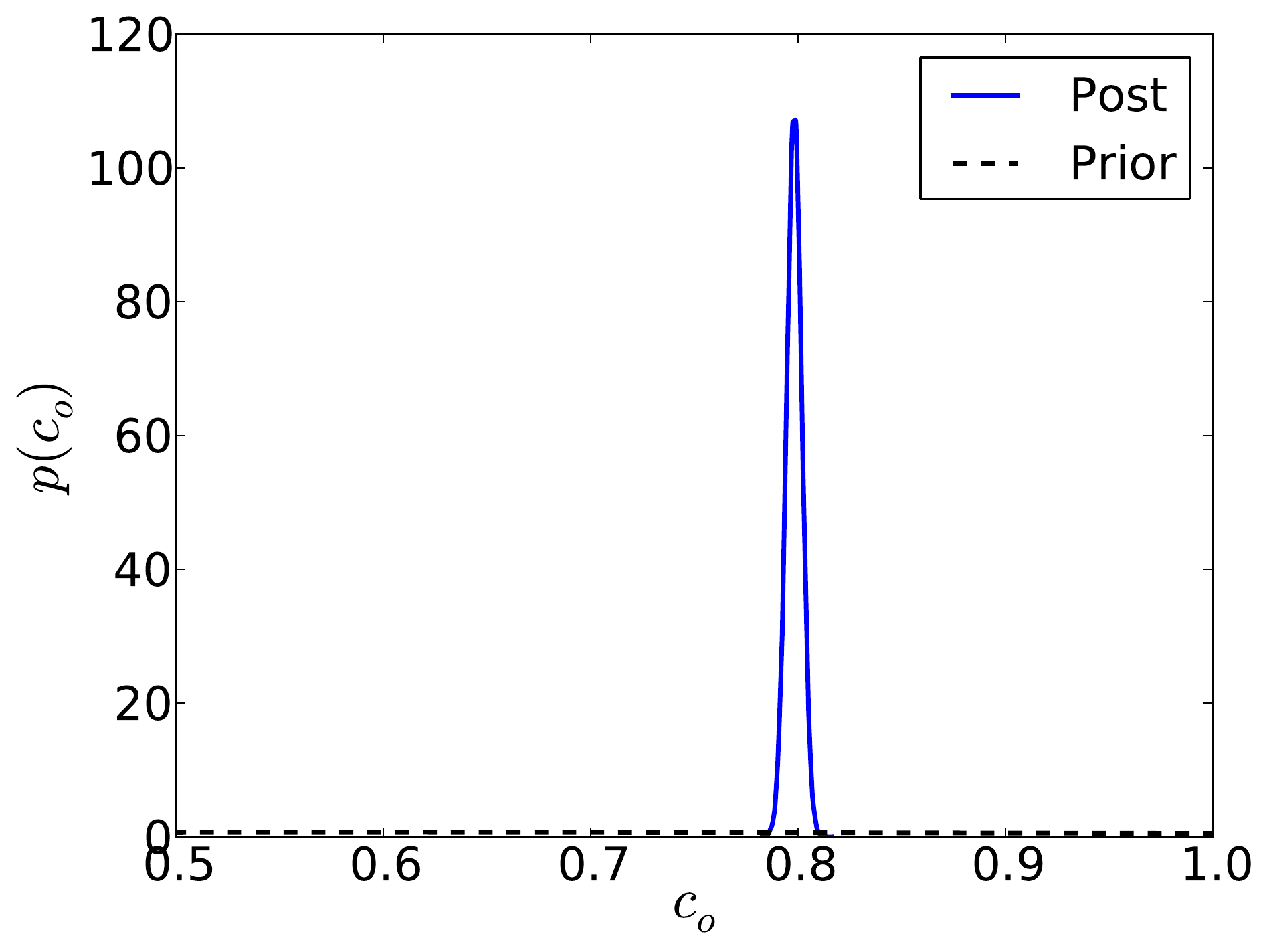}}
\end{center}
\caption{Marginal distributions for parameters $k$ and $c$.  Shown are
 the posterior PDF resulting from a Bayesian
 calibration using the $m=1$ data in Table~\ref{tbl:data_m1} (solid), and the
  prior (dashed).}
\label{fig:si0_params}
\end{figure}

Figure~\ref{fig:si0_params} shows results of calibrating the values of
$k$ and $c$ (with no model inadequacy representation) using Bayesian
inference and the data for the $m=1$ case given in
Table~\ref{tbl:data_m1}.
Clearly, the marginal posterior PDFs are very narrow, indicating that $k$ and
$c$ are highly constrained by the data.  The maximum a posteriori
(MAP) value of $k$ is roughly $2.95$, which is close the true value of
$3.0$.  The MAP value of $c$ is approximately 0.8.  There is no
``true'' value of $c$ since the true damping coefficient is changing
in time.  However, this value is within the range of true values
observed for the actual system for $m=1$.

\begin{figure}[htp]
\begin{center}
\begin{tabular}{cc}
\subfloat[All calibration data.]{
  \includegraphics[width=0.45\linewidth]{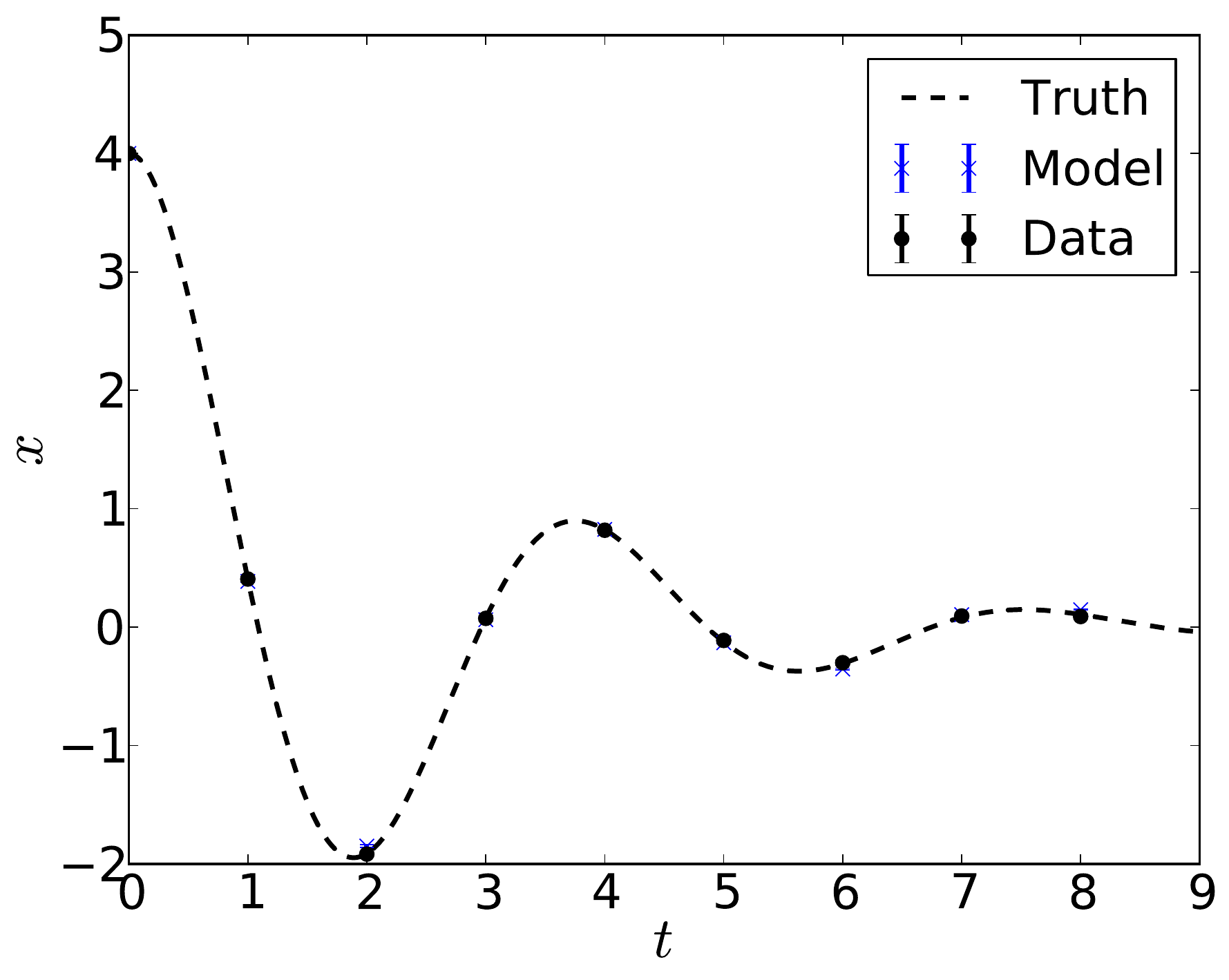}
  \label{fig:si0_cal_challenge_a}} &
\subfloat[HPRD credibility $\gamma$ at $x_{\mathrm{obs}}$ (\ref{eq:gammaHPRD})]{
  \includegraphics[width=0.45\linewidth]{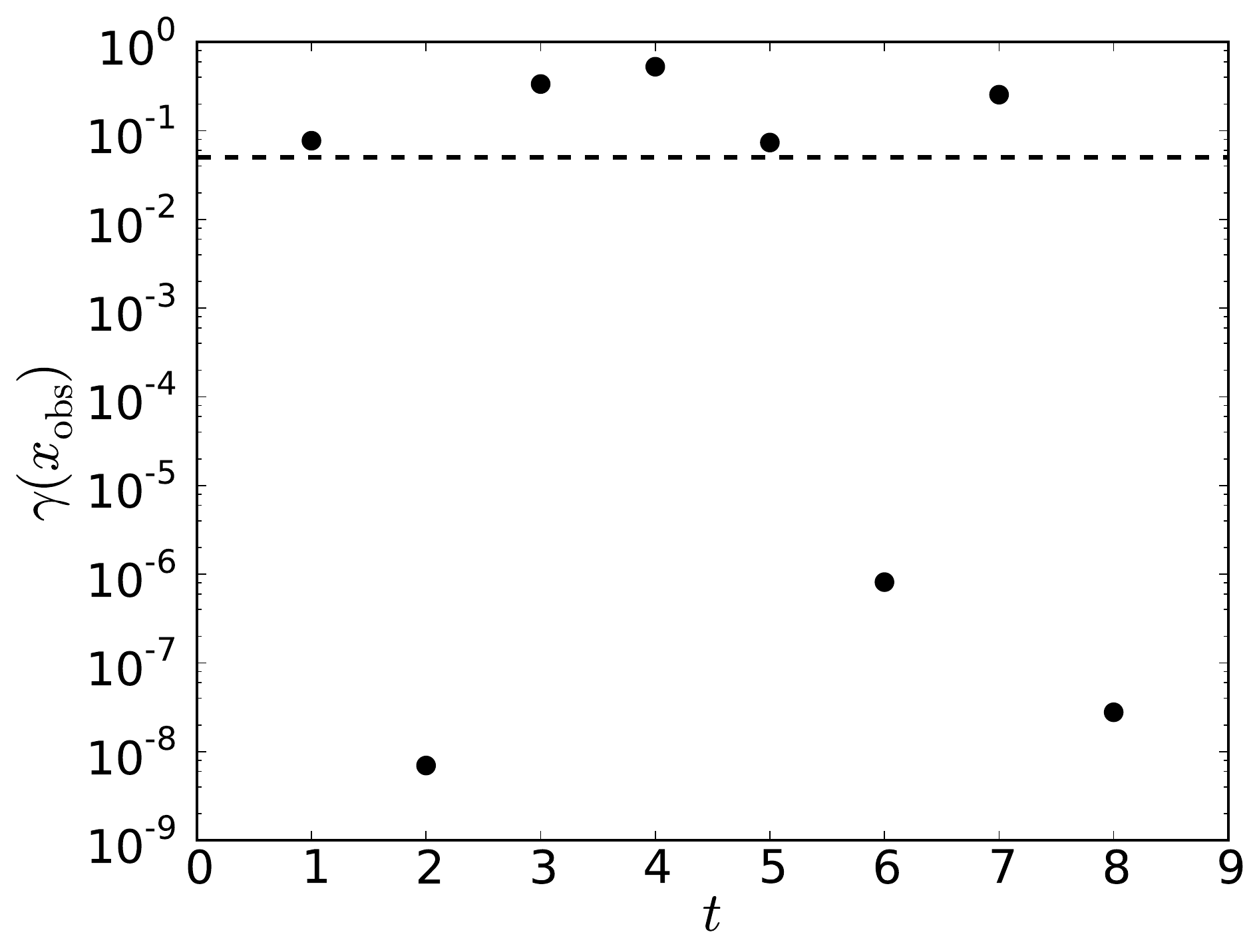}
  \label{fig:si0_cal_challenge_b}} \\
\subfloat[$t=2.$]{
  \includegraphics[width=0.45\linewidth]{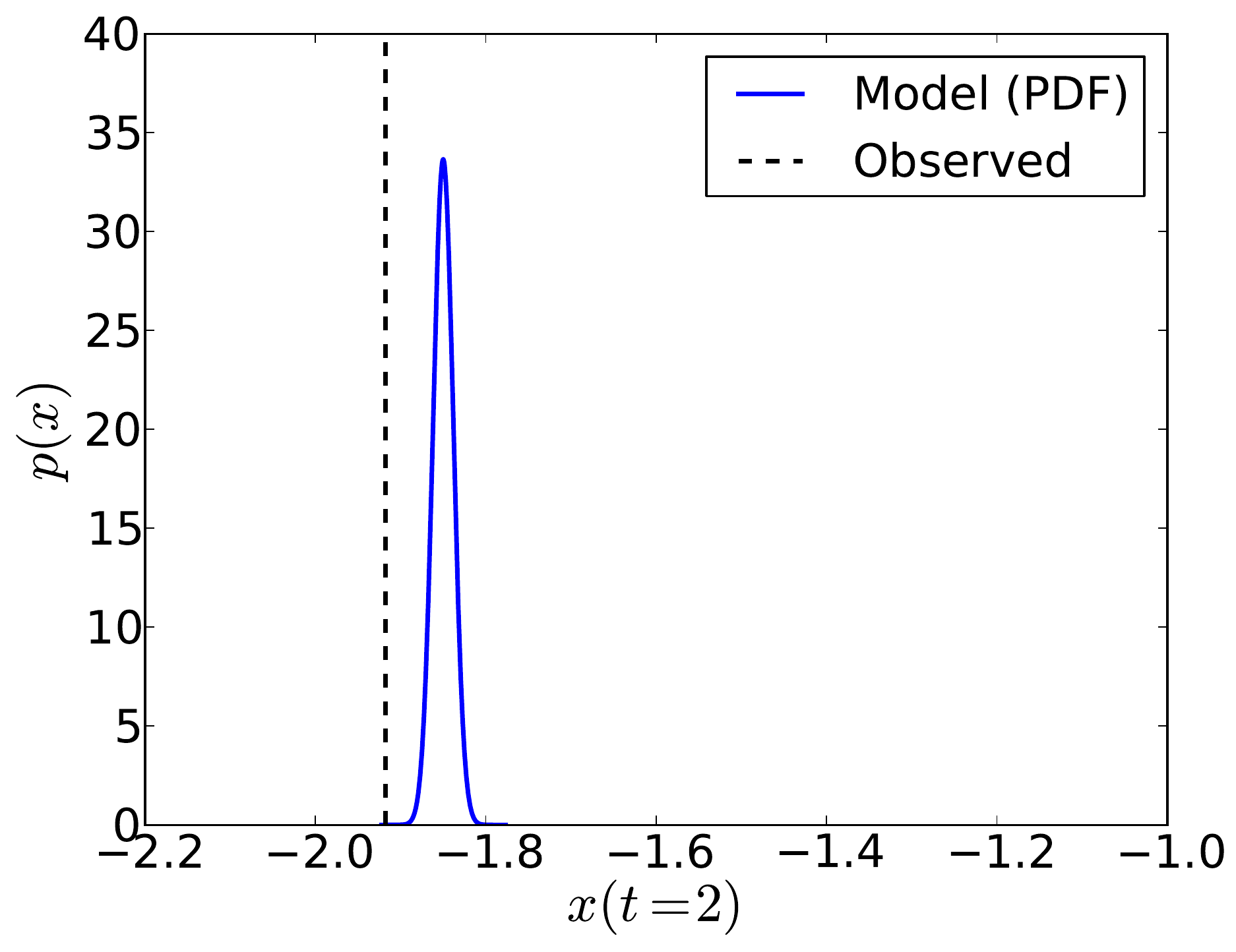}
  \label{fig:si0_cal_challenge_c}} &
\subfloat[$t=7.$]{
  \includegraphics[width=0.43\linewidth]{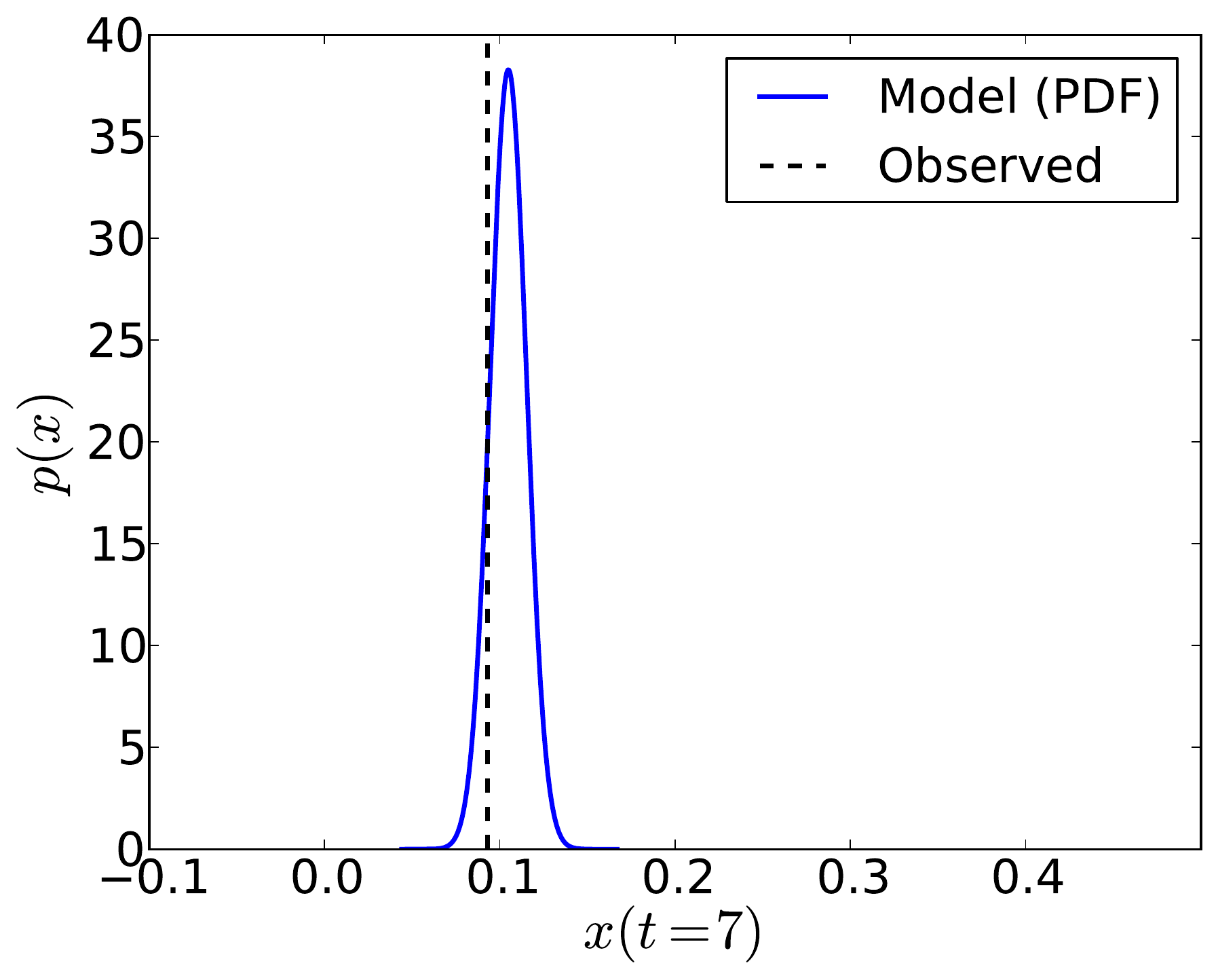}
  \label{fig:si0_cal_challenge_d}}
\end{tabular}
\end{center}
\caption{Comparison of output of calibrated model and observations for $m=1$ for SI0.}
\label{fig:si0_cal_challenge}
\end{figure}

The first validation challenge is to simply compare the output of the
calibrated model with the calibration data, as shown in
Figure~\ref{fig:si0_cal_challenge}. 
%
The figure shows the comparison in a number of different
ways. Figure~\ref{fig:si0_cal_challenge_a} shows the interval
corresponding to the $5^{th}$ and $95^{th}$ percentiles according to
both the model prediction and the observation plus its uncertainty.
However, the errors are quite small relative to the largest values of
$x$, making it difficult to assess the results.
Figure~\ref{fig:si0_cal_challenge_b} shows the HPRD credibility $\gamma$
(\ref{eq:gammaHPRD}), based on the
distribution given by the model plus the observational uncertainty,
and measured relative to a uniform distribution.
To be clear, the model of the observation is given by
\begin{equation*}
x_{\mathrm{obs}} = x_m + \epsilon,
\end{equation*}
where the distribution for $x_m$ is given by forward propagation of
the joint posterior distribution for $k$ and $c$, and $\epsilon \sim
N(0, 0.01)$ is the observational noise.
Figure~\ref{fig:si0_cal_challenge_b} shows that for several of the
observed data points, particularly those at $t=2$, $6$, and $8$, the
corresponding $\gamma$ is less than or equal to approximately
$10^{-6}$, indicating that the actual observation is far out on the
tail of the prediction distribution.  This can be clearly seen in
Figure~\ref{fig:si0_cal_challenge_c}, which shows the prediction
distribution and the observation for $t=2$.  Alternatively, for $t=7$,
where $\gamma$ is significantly larger, there is much better agreement
between the prediction distribution and the observation, as shown in
Figure~\ref{fig:si0_cal_challenge_d}.

However, the existence of points where the actual observation
gives a $\gamma$ that is nearly zero shows that the model developed
based on SI0 does not lead to plausible predictions even for the same
data that was used to calibrate the model.  In this situation, one
must conclude that the uncertainty representation is unable to explain
the differences between the physical model and the data.  This fact
contradicts the modeling assertion that the only important uncertainty
is that due to measurement error.  Since we cannot explain the
observed discrepancies, we cannot confidently extrapolate to the
prediction scenario using this model, even though the actual errors
are not necessarily large.  Despite the fact that the observed errors
are not large, there is no way to characterize what the expected
errors in the prediction scenario might be.  Thus, the model is
invalid for extrapolative prediction.  Note that only the combination
of the physical model and its uncertainty representation have been
invalidated.  It may be possible to improve either to obtain a model
valid for use in the prediction.  For example, we will see that the
same physical model, when equipped with a better uncertainty
description, can make valid predictions.

\subsubsection{State of Information 1}
SI1 does not provide enough information to allow extrapolation,
regardless of the calibration and challenge results.  Specifically,
since we are completely ignorant of the mechanism causing the model
inadequacy, we cannot form a plausible hypothesis 
that would enable
extrapolation.  This is a specific example of a general result,
which is that without concrete hypotheses about the mechanisms causing
the misfit between the model output and the data, extrapolation cannot
be done with confidence.  While the observed misfit can be modeled
statistically, if the misfit cannot be explained,
the effect of the observed errors cannot be extrapolated to the prediction.
The situation is similar to SI0.  We do not have a way to assess the
effect of model inadequacy at the prediction scenario, and thus,
cannot extrapolate with confidence.

\subsubsection{State of Information 2}
For SI2, at least there is enough information to form a hypothesis
that, if correct, enables extrapolation to larger mass.  Specifically,
our hypothesis is that the range of variation of the damping
coefficient $c$ decreases as mass increases.  This hypothesis results
from the observation that, in the system under study, the rate at which
energy is dissipated by the damper is expected to decrease with
increasing mass.  For example, in the system under study, if the
damping is weak, the average over an oscillation of the dissipation
rate is given by $2c k/m$. Assuming that the temperature of the
damping fluid is governed by a competition between how quickly energy
is added to the fluid (by dissipation from the system) and how quickly
energy is transferred to the surroundings, one would expect less
temperature variation, and hence less variation in the damping
coefficient, if energy is dissipated from the system more slowly.  If
this hypothesis is correct, a prediction made for a mass larger than
that used to calibrate the model will be conservative---i.e., the
predicted uncertainty should be larger than necessary to include the
truth.  Thus, the model can be valid in the sense of providing a
conservative extrapolative prediction.

Of course, this is a qualitative argument, and any number of phenomena
could be present in the real system that would make it invalid.  Thus,
we must assess this hypothesis---i.e., validate it---using
observational data.  After calibration, the validation and predictive
assessment processes described below will focus on assessing this
hypothesis.

\begin{figure}[htp]
\begin{center}
\subfloat[$k$]{\includegraphics[width=0.45\linewidth]{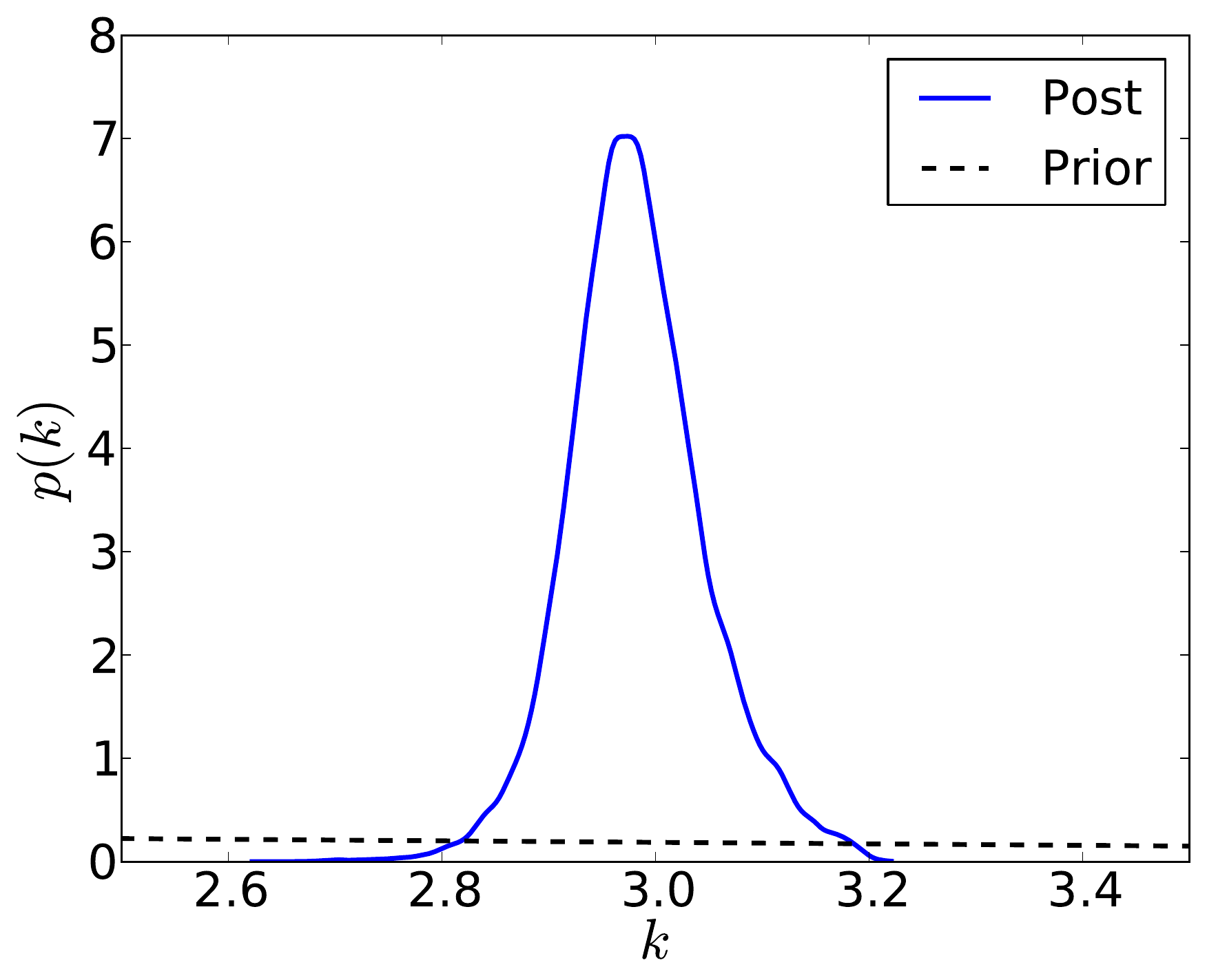}}
\subfloat[$c_{\mu}$]{\includegraphics[width=0.45\linewidth]{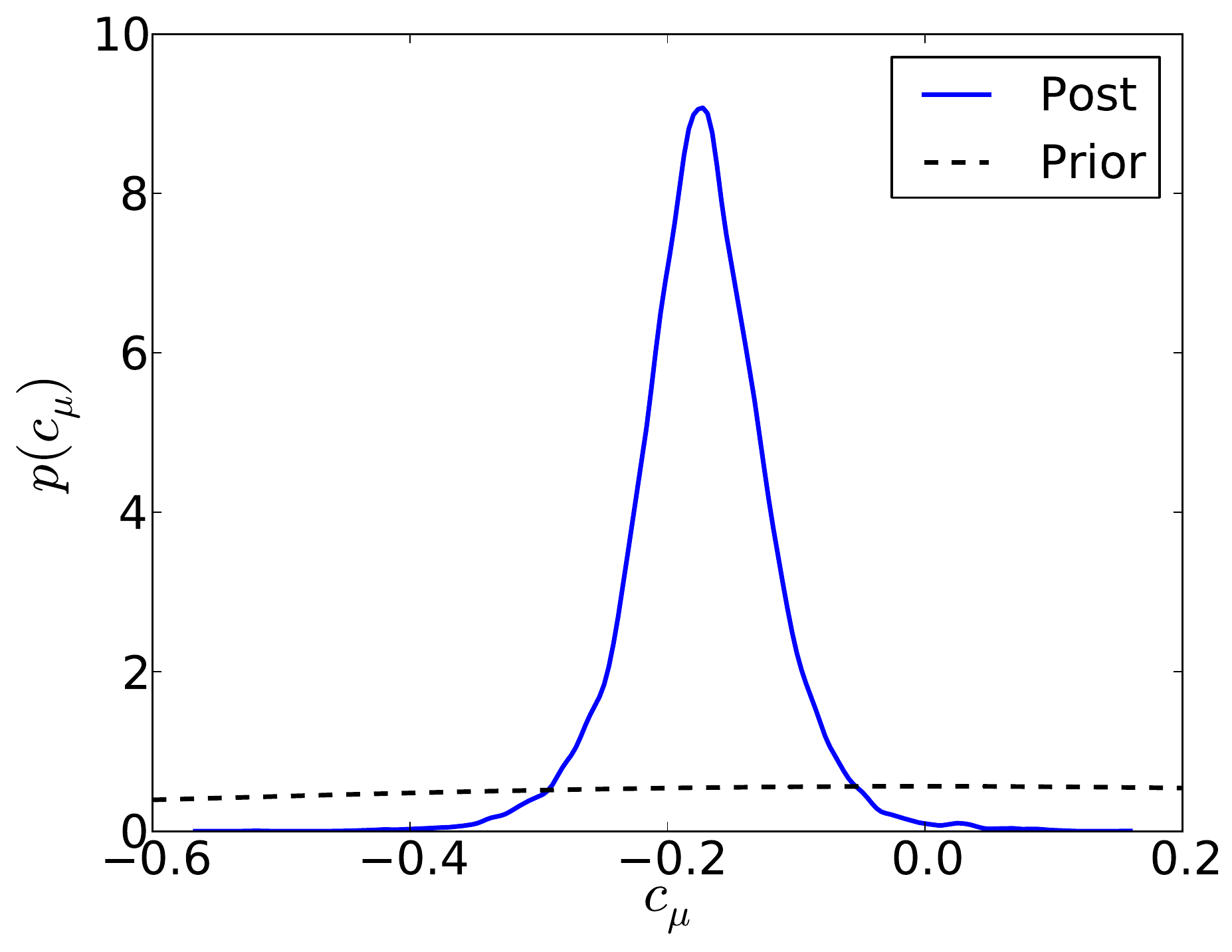}} \\
\subfloat[$c_{\sigma}$]{\includegraphics[width=0.45\linewidth]{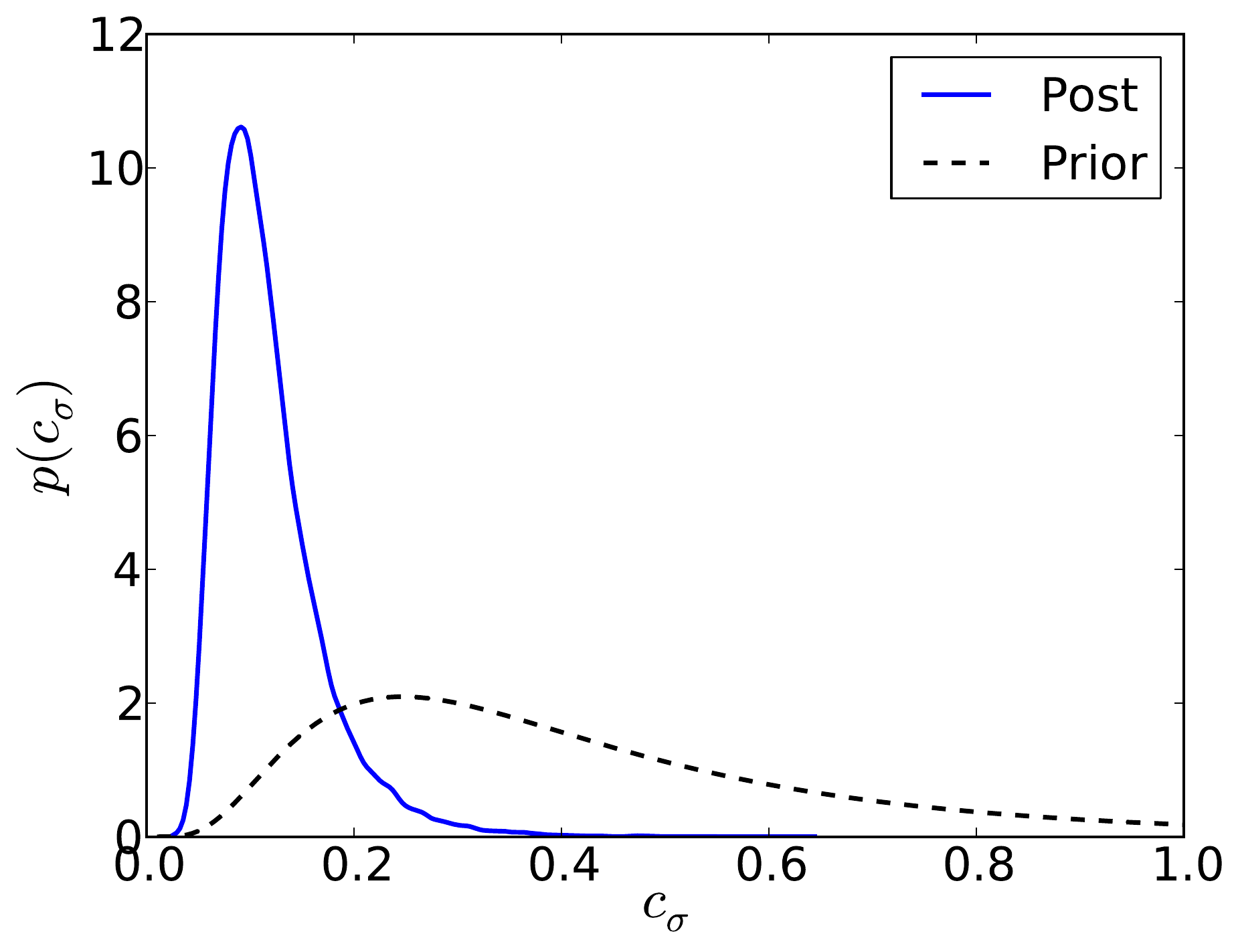}}
\end{center}
\caption{Marginal distributions for parameters $k$, $c_{\mu}$, and
  $c_{\sigma}$.  The solid line labeled ``Post'' shows the marginal
  posterior PDF resulting from a Bayesian update of the prior (dashed
  line labeled ``Prior'') using the $m=1$ data set (Table~\ref{tbl:data_m1}).}
\label{fig:si2_params}
\end{figure}

Figure~\ref{fig:si2_params} shows the marginal PDFs resulting from the
Bayesian calibration using the $m=1$ data.
Note that the marginal posterior for $k$ is broader than in the SI0
result and that the (unknown) true value $k = 3$ is in the support of
the posterior PDF.

In the validation phase, the goal is to check that the model, is
capable of representing both the $m=1$ data, on which it was
calibrated, and the $m=2$ data.  If any data point is highly unlikely
according to the model, it is declared invalid, as happened with the
SI0 model when it failed to reproduce the calibration data.

\begin{figure}[htp]
\begin{center}
\begin{tabular}{cc}
\subfloat[All calibration data.]{
  \includegraphics[width=0.45\linewidth]{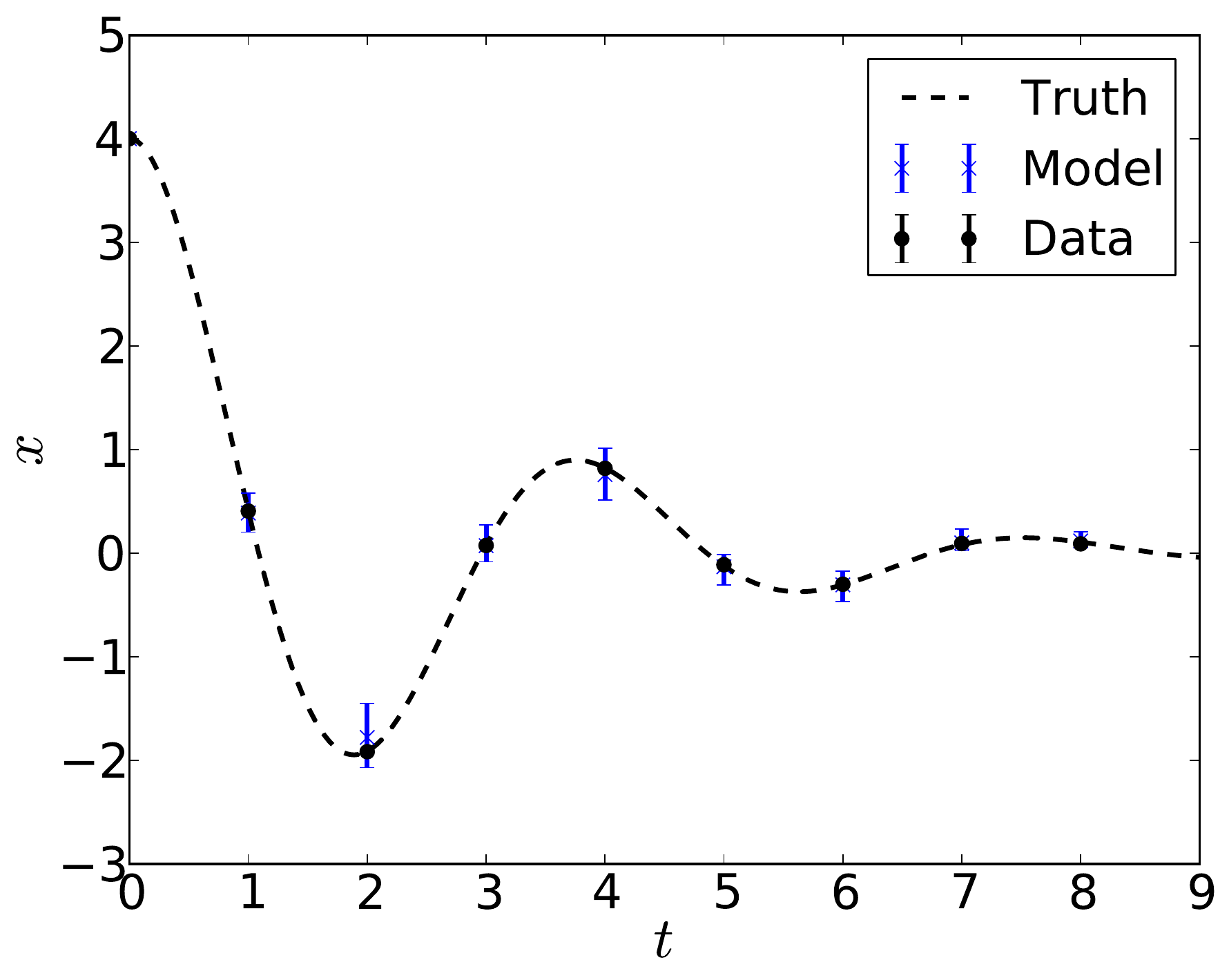}
  \label{fig:si2_cal_challenge_a}} &
\subfloat[HPRD credibility $\gamma$ at $x_{\mathrm{obs}}$ (\ref{eq:gammaHPRD})]{
  \includegraphics[width=0.45\linewidth]{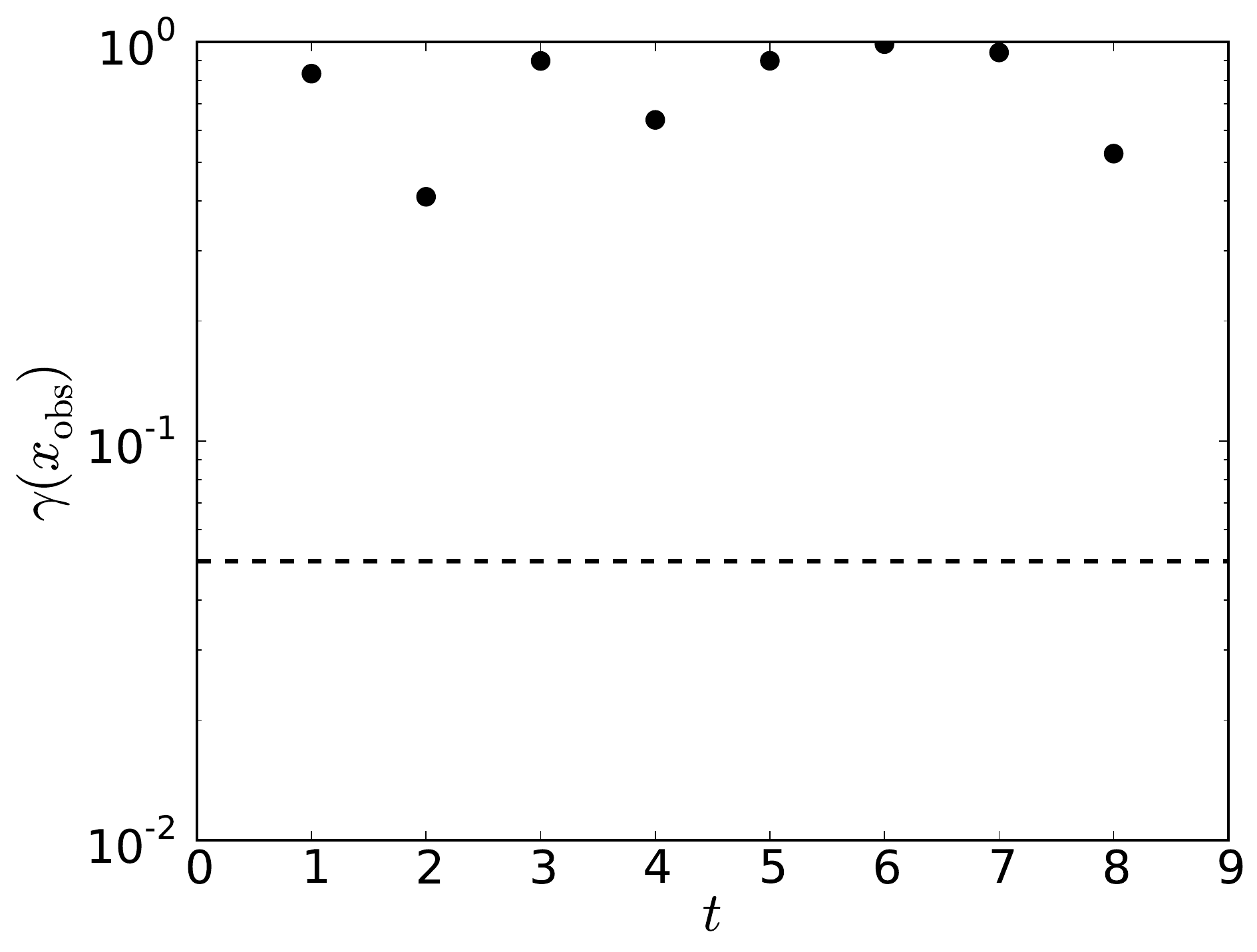}
  \label{fig:si2_cal_challenge_b}} \\
\subfloat[$t=2.$]{
  \includegraphics[width=0.45\linewidth]{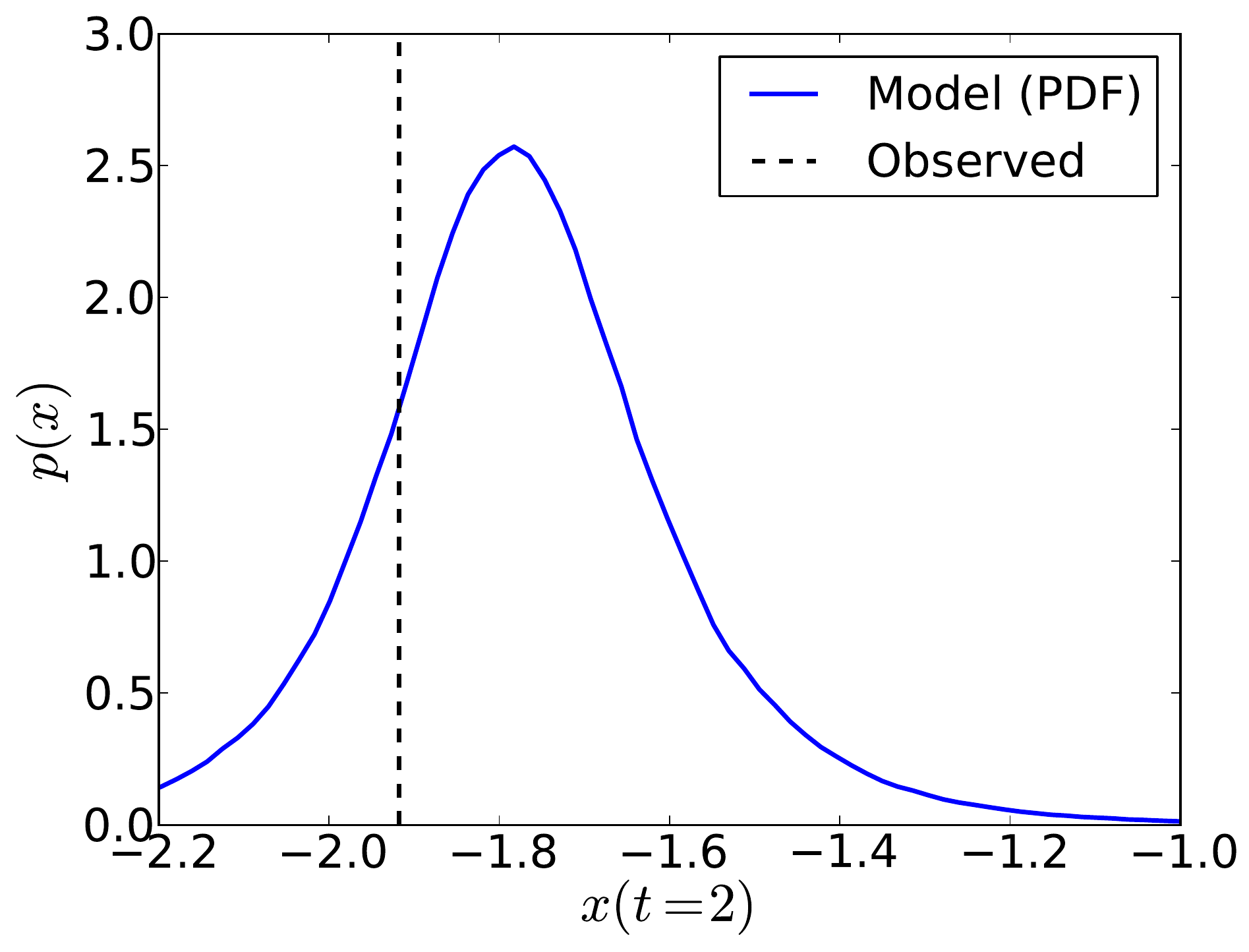}
  \label{fig:si2_cal_challenge_c}} &
\subfloat[$t=7.$]{
  \includegraphics[width=0.43\linewidth]{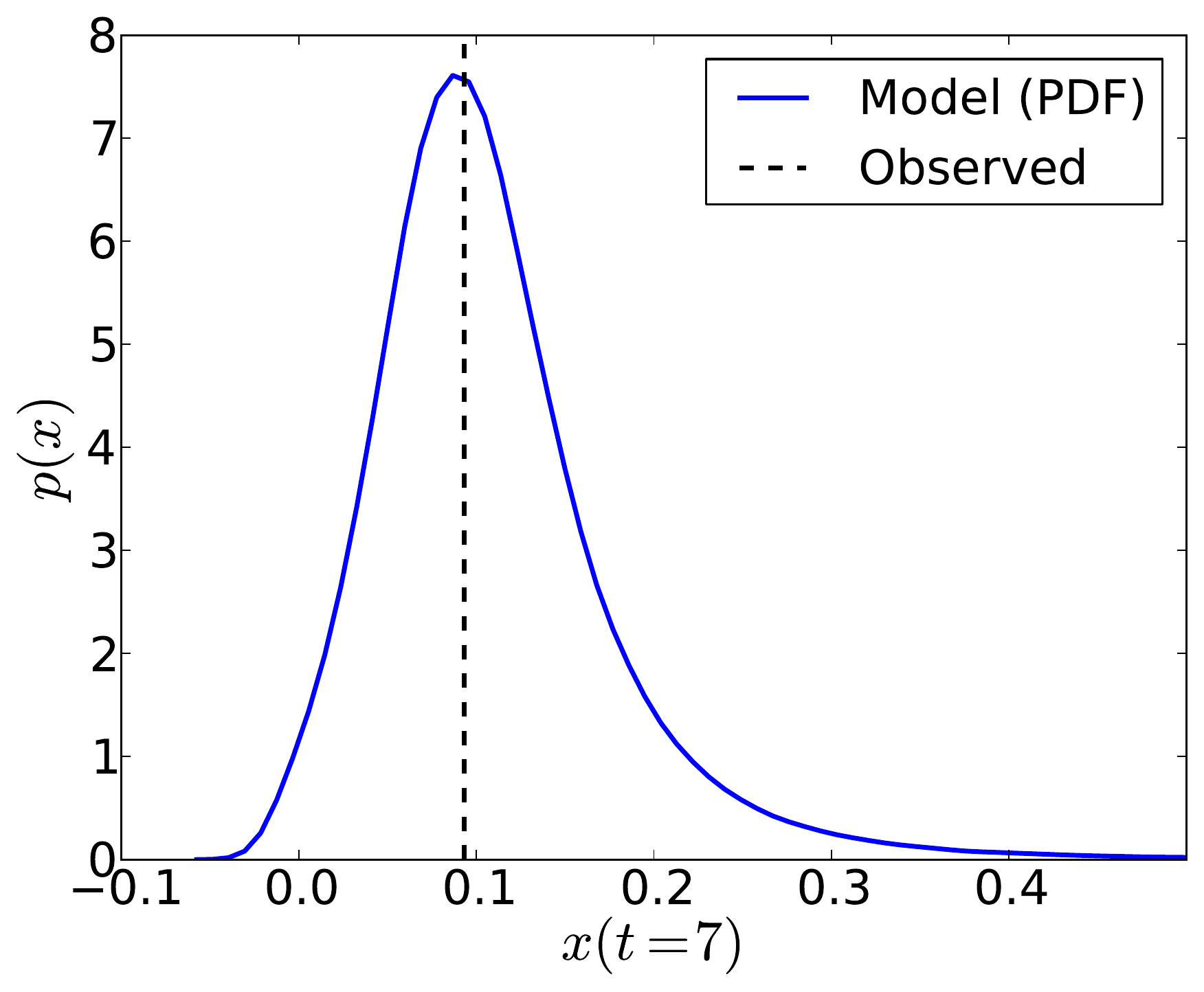}
  \label{fig:si2_cal_challenge_d}}
\end{tabular}
\end{center}
\caption{Comparison of output of calibrated model and observations for $m=1$ for SI2.}
\label{fig:si2_cal_challenge}
\end{figure}

\begin{figure}[htp]
\begin{center}
\begin{tabular}{cc}
\subfloat[All calibration data.]{
  \includegraphics[width=0.45\linewidth]{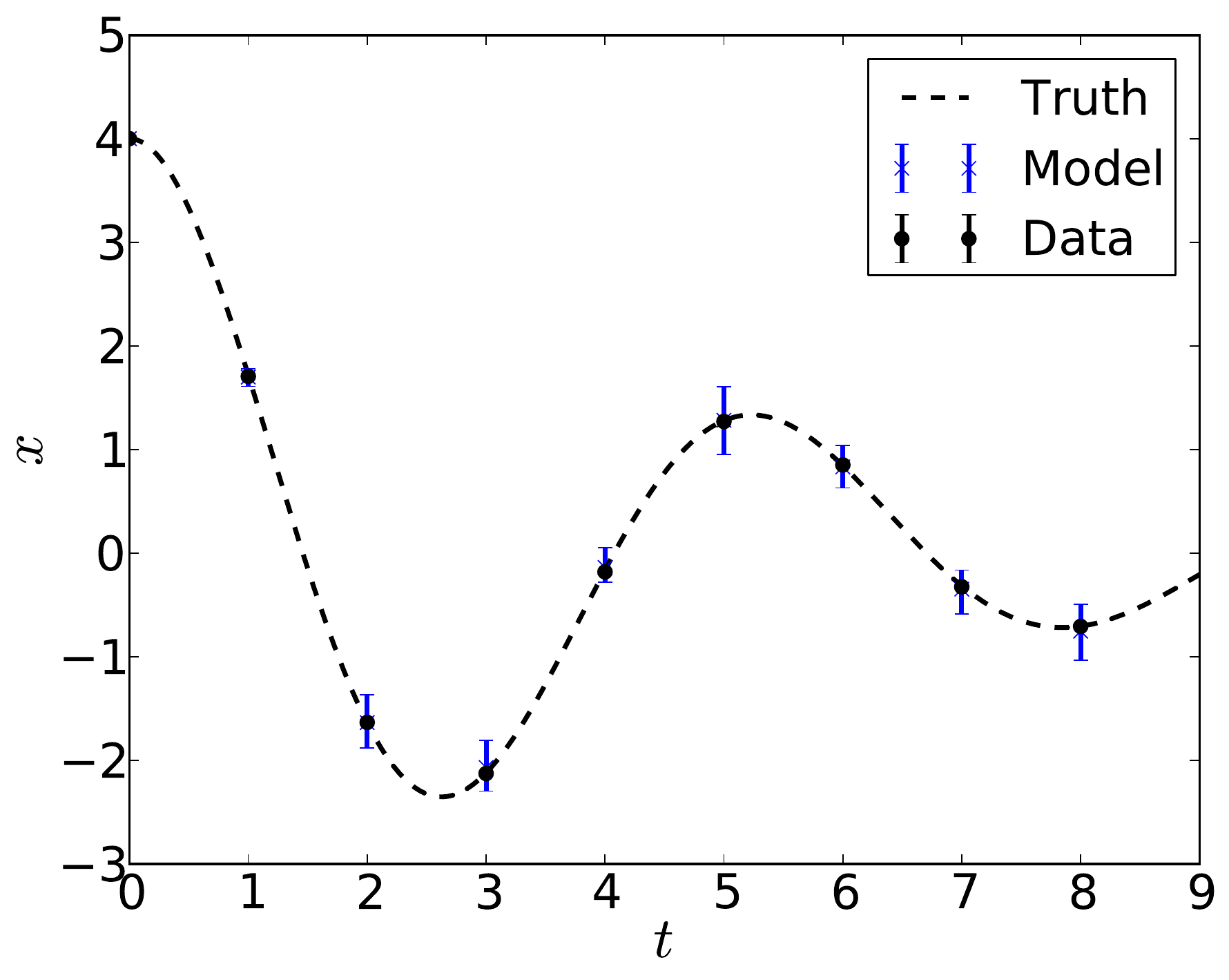}
  \label{fig:si2_val_challenge_a}} &
\subfloat[HPRD credibility $\gamma$ at $x_{\mathrm{obs}}$ (\ref{eq:gammaHPRD})]{
  \includegraphics[width=0.45\linewidth]{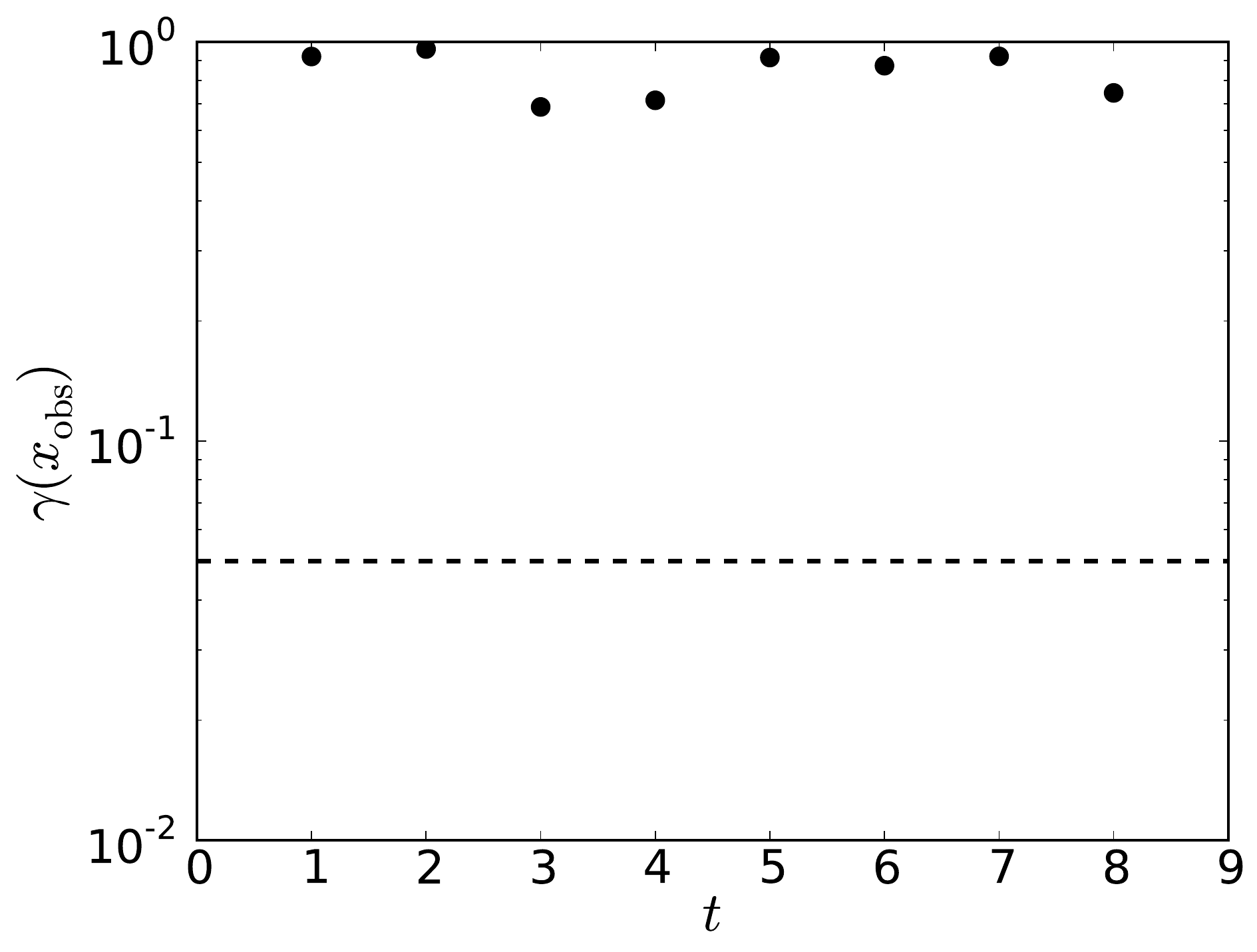}
  \label{fig:si2_val_challenge_b}} \\
\subfloat[$t=2.$]{
  \includegraphics[width=0.45\linewidth]{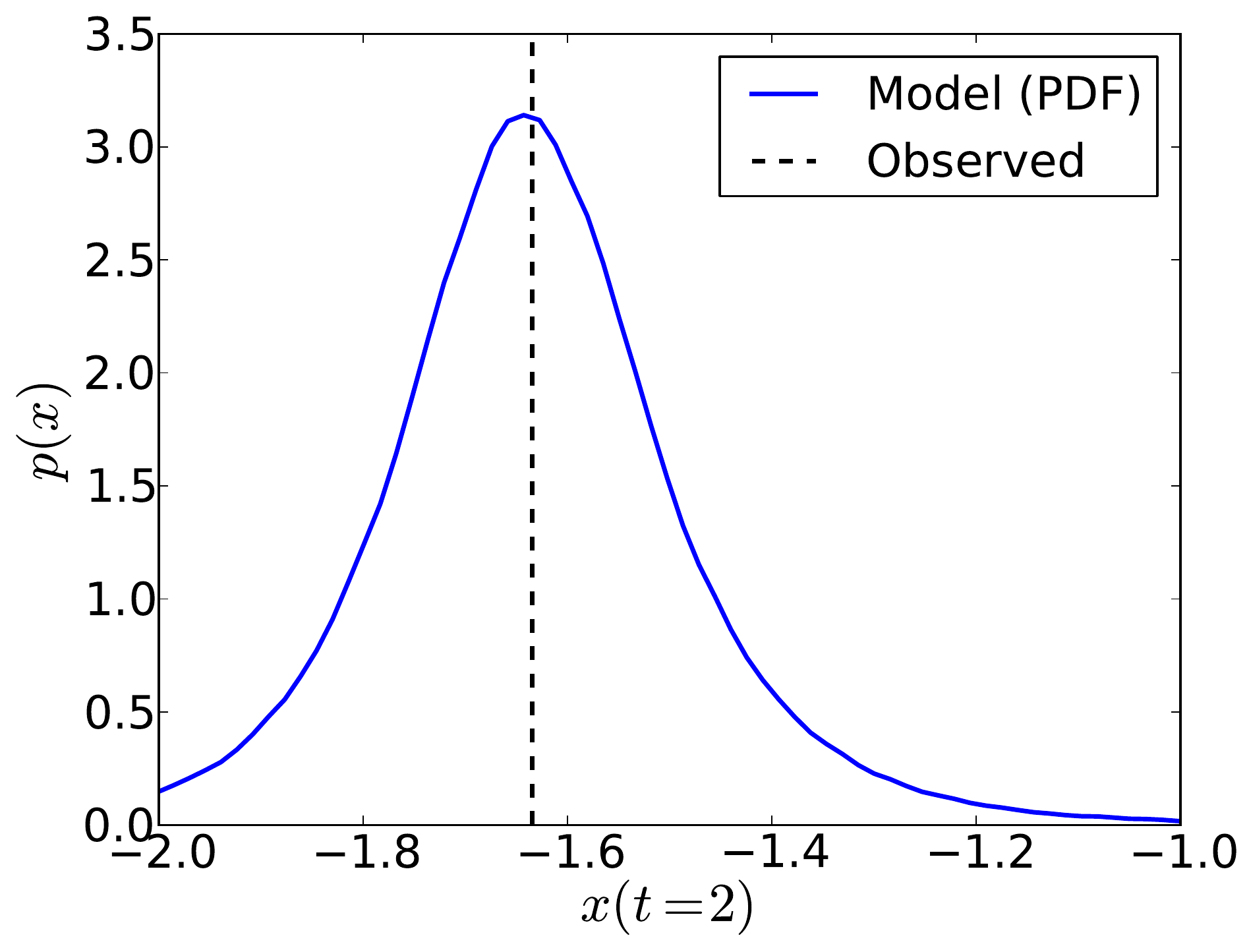}
  \label{fig:si2_val_challenge_c}} &
\subfloat[$t=7.$]{
  \includegraphics[width=0.43\linewidth]{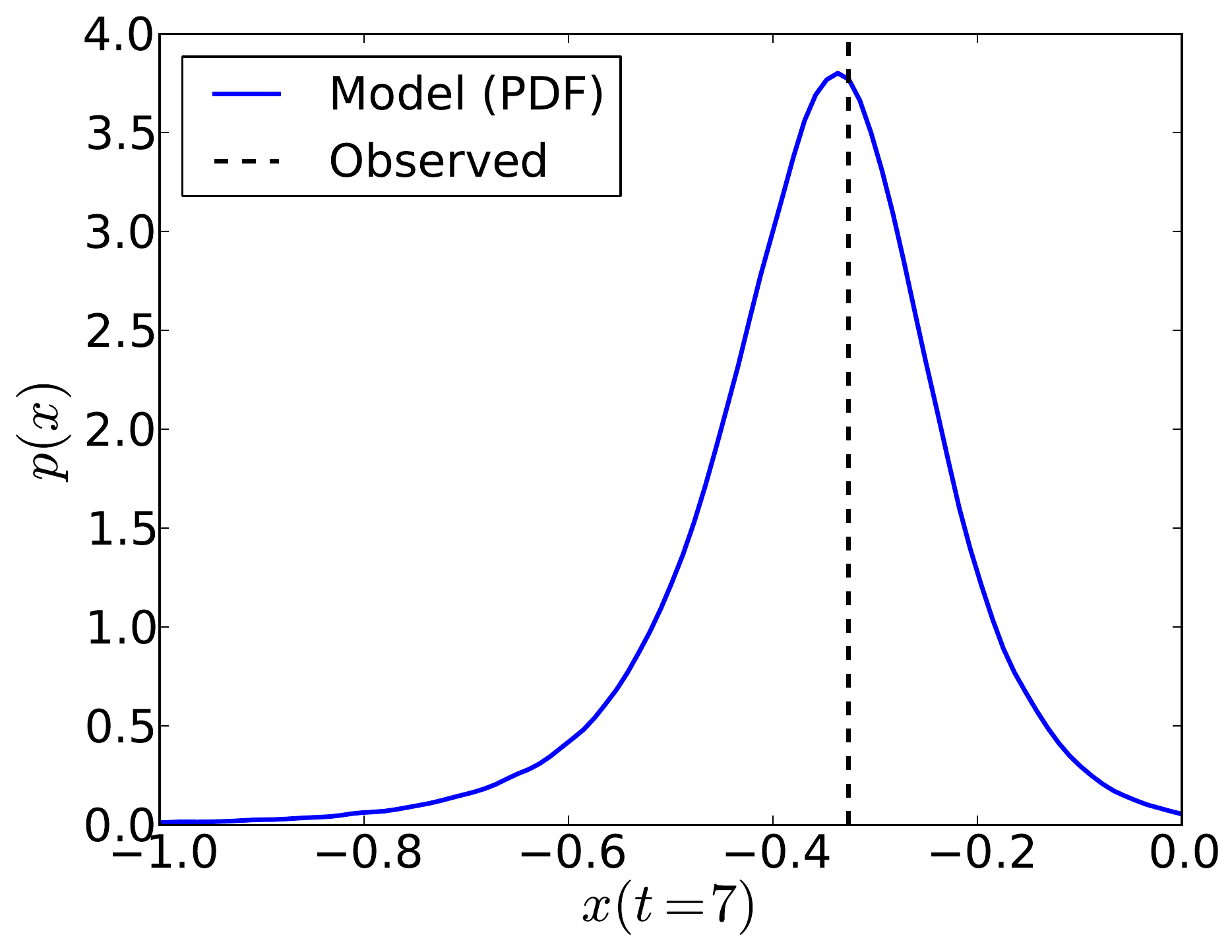}
  \label{fig:si2_val_challenge_d}}
\end{tabular}
\end{center}
\caption{Comparison of output of calibrated model and observations for $m=2$ for SI2.}
\label{fig:si2_val_challenge}
\end{figure}

Figure~\ref{fig:si2_cal_challenge} shows a comparison between the
calibration data and the output of the calibrated model.  The
quantities shown in Figure~\ref{fig:si2_cal_challenge} are analogous
to those shown for SI0 in Figure~\ref{fig:si0_cal_challenge}, but
unlike SI0, the predictions given are much more uncertain and agree
better with the observations.  In particular, the
HPRD credibility, $\gamma$,
shown in~\ref{fig:si2_cal_challenge_b} are never less than
approximately 0.4, indicating that the observations are near the
highest probability density regions of the prediction distribution, as
shown in Figures~\ref{fig:si2_cal_challenge_c}
and~\ref{fig:si2_cal_challenge_d}.
The same statement holds for the $m=2$ data, which was not used in
calibration, as shown in Figure~\ref{fig:si2_val_challenge}.

Since there is no evidence to invalidate the model after comparing the
calibrated model predictions with all the available data, the
validation phase is complete, and we may move on to the predictive
assessment.  To begin, recall that the predictive assessment is
necessarily problem specific.  In the current context, a reliable
prediction is dependent on the hypothesis regarding the dependence of
the variability of the damping coefficient on the mass. Thus, as part
of predictive assessment, it is necessary to test this hypothesis to
the fullest extent possible.  Here, we test this hypothesis by
performing a separate model calibration using the the $m=2$ data set.
If the uncertainty in $c$ required to fit the $m=2$ data is smaller
than that required to fit the $m=1$ data, as measured by the posterior
results, then $c$ must be varied less to match the larger mass data,
and the available data supports the hypothesis.

\begin{figure}[htp]
\begin{center}
\subfloat[$k_s$]{
  \includegraphics[width=0.45\linewidth]{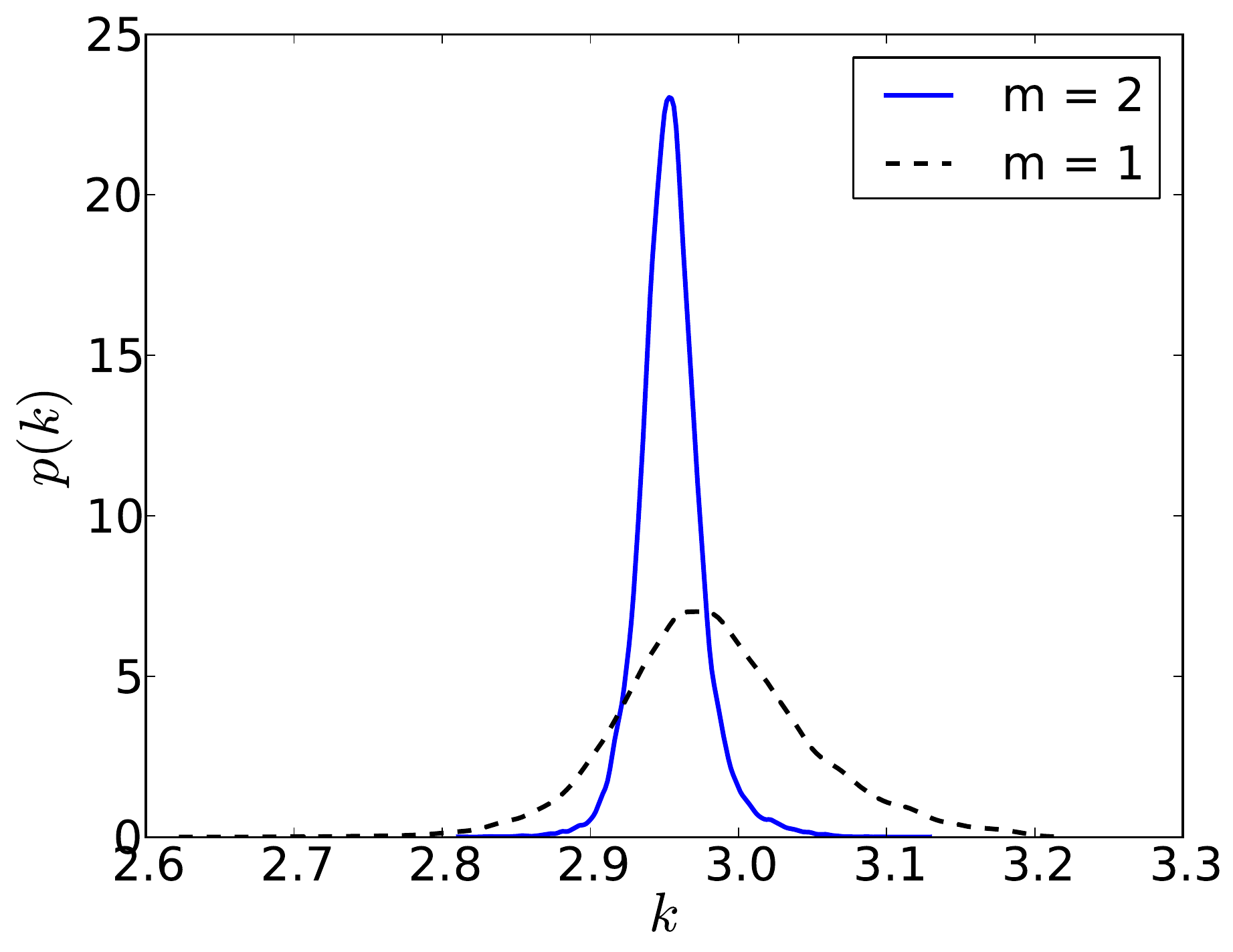}} \\
\subfloat[$c_{\mu}$]{
  \includegraphics[width=0.45\linewidth]{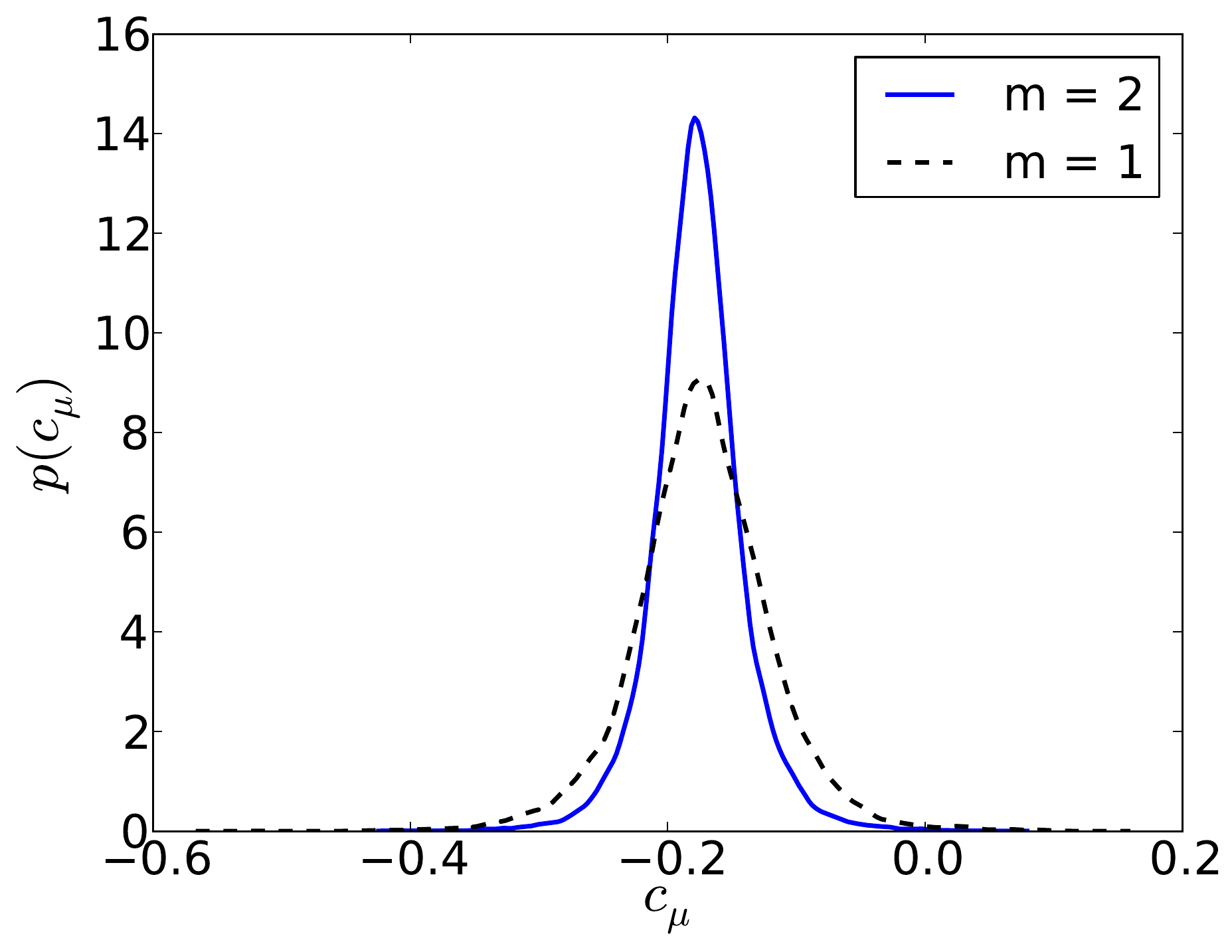}}
\subfloat[$c_{\sigma}$]{
  \includegraphics[width=0.45\linewidth]{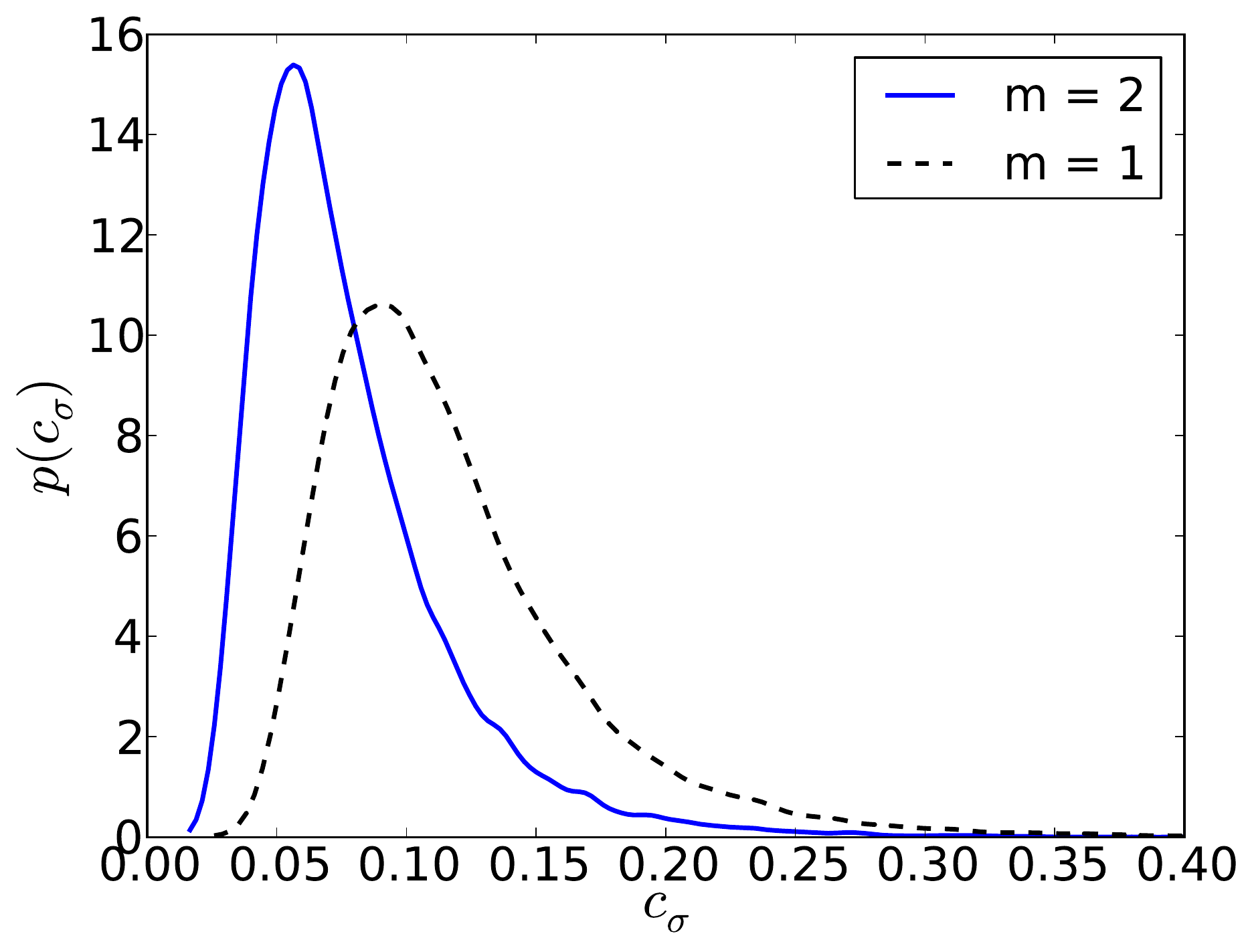}}
\end{center}
\caption{Comparison of marginal posterior PDFs obtained from a
  Bayesian update using the $m=2$ data (solid lines) versus the
  $m=1$ data (dashed lines).}
\label{fig:si2_val_challenge_recalibrate}
\end{figure}

\begin{figure}[htp]
\begin{center}
\subfloat[Maximum likelihood result.]{
  \includegraphics[width=0.45\linewidth]{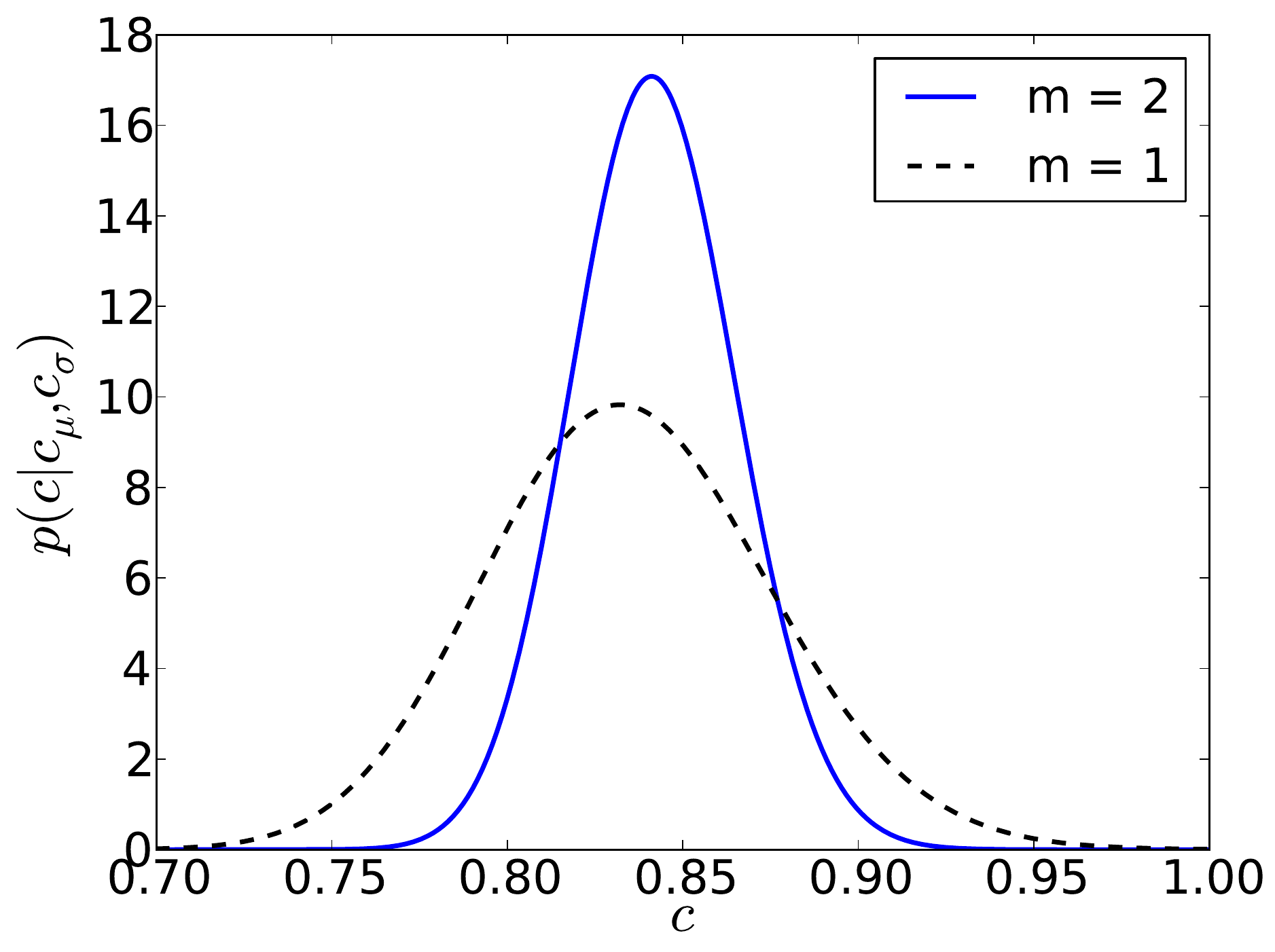}}
\subfloat[Standard deviation of $c$.]{
  \includegraphics[width=0.45\linewidth]{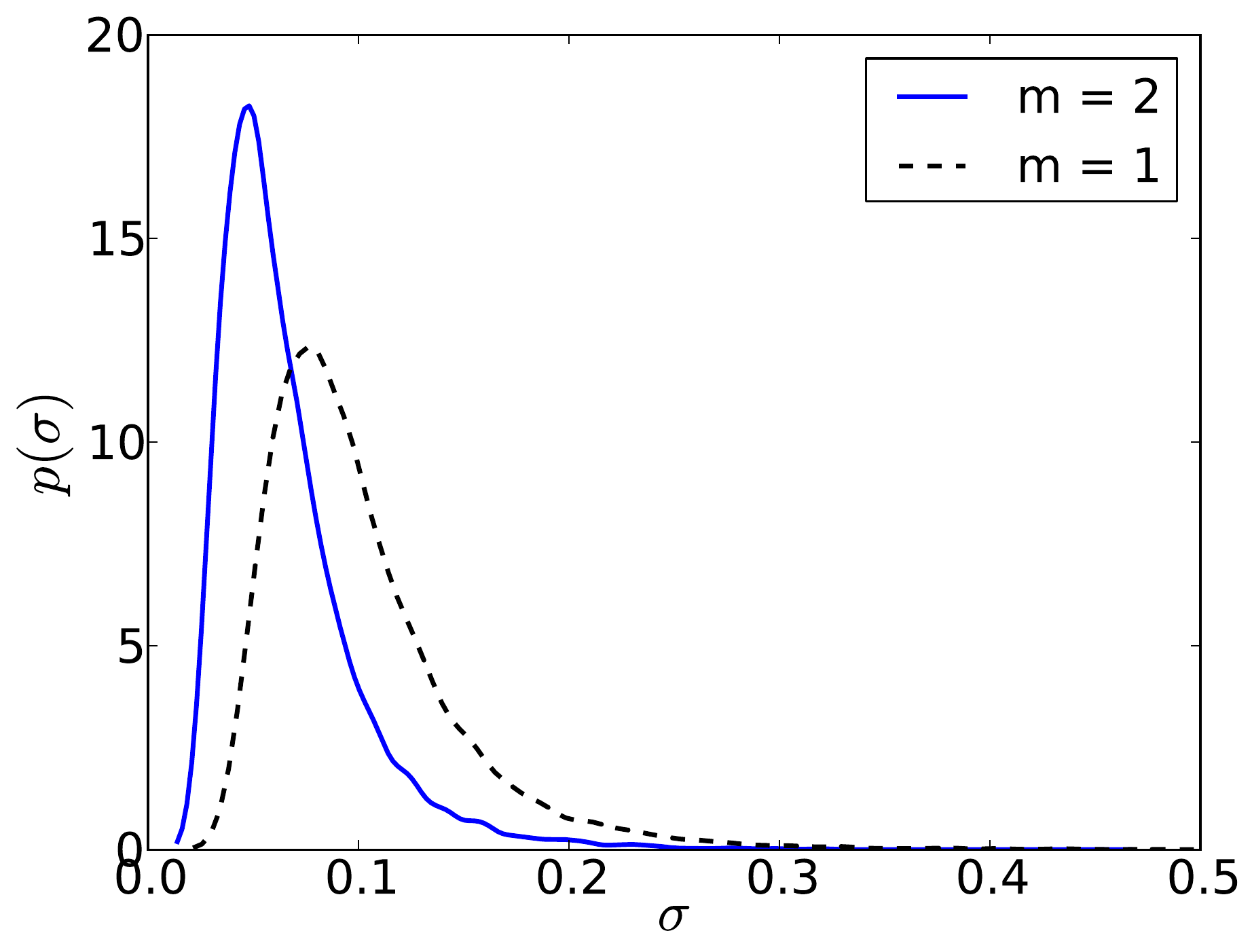}}
\end{center}
\caption{Comparison of maximum likelihood log-normal distribution of
  $c$ and the PDF for the standard deviation of $c$ for the model
  calibrated using both the $m=1$ (dashed) and $m=2$ (solid)
  data.}
\label{fig:si2_val_challenge_implied_c}
\end{figure}

Figure~\ref{fig:si2_val_challenge_recalibrate} shows the results of
this test.
The marginal posterior PDFs for $k$ and $c_{\mu}$ based on the two data
sets are largely
consistent, although the $m=2$ data somewhat better informs the
parameters.  However, $c_{\sigma}$, which is the standard
deviation of $\log(c)$, moves significantly to the left.  This result
implies that the standard deviation of $c$ decreases.  This fact is
demonstrated in Figure~\ref{fig:si2_val_challenge_implied_c}, which shows the
distribution of $c$ corresponding to the maximum likelihood values
of $c_{\mu}$ and $c_{\sigma}$ as well as the PDF for the standard
deviation of $c$ implied by the posterior joint distribution of
$c_{\mu}$ and $c_{\sigma}$.
Clearly the maximum likelihood distribution of $c$ is narrower when
using the data from the larger mass.  In particular, the standard deviation of
the maximum likelihood model decreases from approximately $0.041$ to
$0.023$, a decrease of nearly $44\%$.  The distribution for $\sigma$
(the standard deviation of $c$) shows a similar result.  When using
the $m=2$ data, the probability distribution is shifted to lower values than
when using the $m=1$ data, indicating that the variability of
$c$ required to fit the data is decreasing with $m$.  This result
is consistent with the hypothesis.

Given this result, we can move to additional predictive assessments.
In general, there may be many additional hypotheses to test or
sensitivities to check, as discussed in \S\ref{sec:predict_assess}.
In this case we note that the data are quite informative about the
parameters of the SI2 model (see Figure~\ref{fig:si2_params}),
indicating that there are no uninformed aspects of the model to which
the predictions could be sensitive.  Further, the ``domain of
applicability'' of the model is implicitly checked (to the extent
possible with the data) by the results of the calibration with the
$m=2$ described previously.  Thus, we conclude that there is good
reason to trust the calibrated SI2 model to make credible predictions
for the $m=5$ case.

\begin{figure}[htp]
\centering \includegraphics[width=0.6\linewidth]{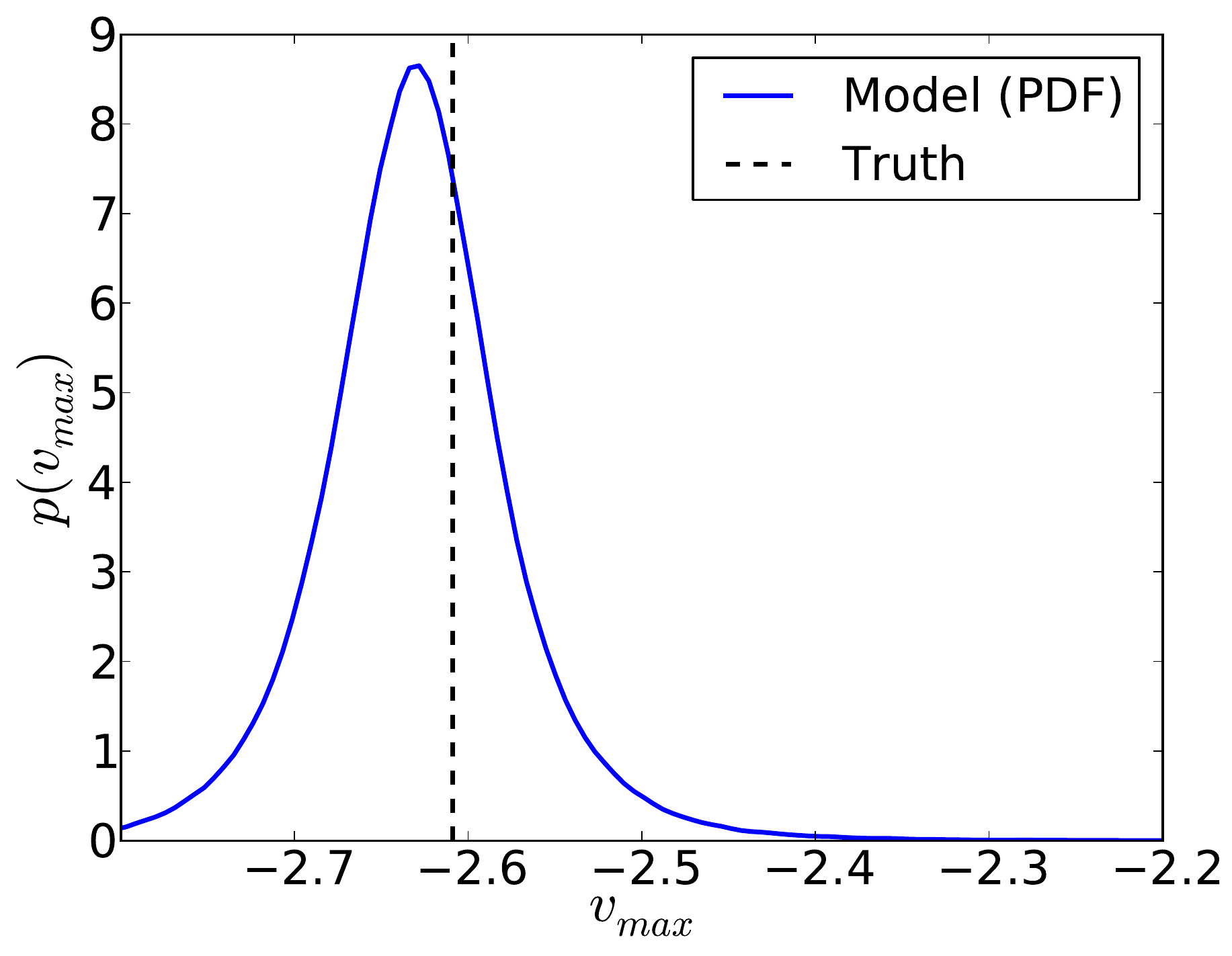}
\caption{Prediction of maximum velocity for $m=5$.}
\label{fig:qoi}
\end{figure}

Having challenged the model and the hypothesis needed to make
extrapolation to larger masses possible, we are ready to make the
prediction.  Recall that the QoI is the maximum velocity for $m=5$.
Figure~\ref{fig:qoi} shows the prediction given by the model as well
as the true value.
Of course, in general the true value is unknown.  It is given here
just to show that the process has led us to the correct
conclusion---i.e., the extrapolation is valid in the sense that the
truth is assigned reasonable likelihood by the model.  

Given this validated prediction, the next step would be to ask whether
the predicted uncertainty is small enough to inform the desired
decision.  Since any tolerance we could specify here would be entirely
contrived and artificial, we choose not to pursue this aspect.
However, this is the simplest aspect of the predictive validation
framework and should not cause significant difficulty.  If the
uncertainty is deemed too large, one would have to pose a better
physical model, pose better uncertainty representations, get more
data, or some combination of these, and begin the validation process
again.

\section{Conclusions} 
\label{sec:conclusions}
The predictive validation process proposed here provides a framework
for building confidence in extrapolative predictions issued by models
based on physics.  There are two key ingredients enabling reliable
extrapolation with such models.  First, it is common that such models
are based upon highly-reliable theory that is augmented with
less-reliable embedded models to form a composite model.  Second, the
scenario dependence of the embedded models is generally different from
the scenario of the full composite model, allowing the full model to
be used for extrapolation without extrapolating the lower fidelity
aspects.  Given these ingredients, the predictive validation process
requires the specification of a model inadequacy representation for
the low-fidelity embedded models.  This representation enables one to
connect the QoI with observational data in a way that previous
approaches lack.  Once a physical model and inadequacy representation
have been specified, the model is subjected to a calibration phase,
where the observations are used to inform uncertain aspects of the
model.  Then, in validation, the model is challenged with new
observational data.  The primary question in this validation test is
whether the observations are plausible given the uncertainty in the
prediction.  That is, the main objective of the validation is to test
the combination of physical and uncertainty models.  Finally, if
the validation is satisfactory, the model is subjected to a predictive
assessment to determine if the predictions of the QoIs can be trusted
and whether they are sufficient from the point of view of a decision
maker.

The full process has been illustrated using a simple
spring-mass-damper system where the true physics of the damper are not
well-understood by the modeler.  Despite this lack of knowledge, the
process is able to build confidence in a very simple model.  However,
this gain in confidence is contingent on the information available
at the beginning of the process.  With some specific knowledge
regarding the cause of the modeling error, one may be able to build
confidence in an inadequate model for making a particular prediction.
Without this kind of information---i.e., simply knowing that the model
does not match reality and nothing more---one will generally be able
to do very little.

While the global process presented here is clear, there are several
research and development issues that need to be addressed to enable
the validation of such predictions in a wide range of
applications. These research challenges are outlined briefly below.
\begin{enumerate}
\item {\bf Inadequacy models:} A critical component of the proposed
      process is a probabilistic model of the errors in the embedded
      models. Such an inadequacy model should respect all that is known
      about the approximations and deficiencies of the models, all that
      is known about the quantities being modeled, and the available
      data. Broadly applicable techniques for formulating these
      inadequacy models are needed, especially for situations where the
      modeled quantity is a field.
\item {\bf Data uncertainty models:} The uncertainty in experimental
      data is a critical input to the process, and better
      characterizations of this uncertainty are needed. Of particular
      concern are characterizing dependencies among different data
      points and uncertainties arising from data reduction modeling.
\item {\bf Representing qualitative information:} In Bayesian analysis,
      posing priors that faithfully represent what is known about the
      problem at hand is important to making reliable inference. Once
      the prior knowledge is expressed mathematically, one often has 
      rigorous tools, e.g., maximum entropy, to construct the needed prior 
      distribution.  But, this knowledge is
      commonly qualitative and difficult to express
      mathematically. Tools and techniques are needed to formulate the
      kinds of qualitative knowledge we commonly have regarding physical models
      based on reliable theory, as discussed here. Representations of
      qualitative information are also important in characterizing
      modeling inadequacy and data uncertainty.
\item {\bf Domains of applicability:} It is critical to predictive
      validation to identify when an embedded model is being used under
      conditions for which it has not been calibrated and validated. For
      many models, an appropriate set of model-specific scenario
      parameters has not been defined. Determining such scenario
      parameters is part of physical modeling, and therefore dependent on
      the phenomena being modeled, and it is in general a significant
      challenge. However, generally applicable tools and techniques for
      developing and evaluating such parameterizations are needed.
\item {\bf Experimental design:} Data is needed for calibration and
      validation, but it is critical to have data that adequately informs
      the QoIs in the context of the models. That is, measurements of
      quantities that are sensitive to the same uncertainties as the QoIs
      are needed under scenarios that will produce a sufficiently large
      domain of applicability for the embedded models. Metrics are needed
      to rank potential validation cases, allowing the best experimental
      measurements and scenarios to be determined automatically.
\item {\bf Computational algorithms:} While we have not discussed the
      computational tools needed to execute the predictive validation
      process discussed here, there are significant algorithmic
      challenges associated with high dimensional probability spaces
      (the curse of dimensionality), with expensive computational
      models and with stochastic models, which arise naturally from the
      inadequacy representations discussed here.
\end{enumerate}
As should be clear in the above discussion, research challenges 1-4
essentially require introducing knowledge about the physical phenomena
being modeled into the process. Advancing techniques to address these
challenges will presumably require pursuing them in the context of
a variety of specific physical systems. 

\section*{Acknowledgments}
This material is based on work supported by the Department of 
Energy [National Nuclear Security Administration] under Award Number
[DE-FC52-08NA28615]. The authors are grateful to Profs. Tinsley Oden and
Ivo Babuska, and Dr. David Higdon for many insightful discussions.


\appendix
\section{Truth System}
\label{app:truth}
The truth system is similar to the physical model discussed in
\S\ref{sec:results}.  However, instead of having a constant damping
coefficient, the damping coefficient is a function of the temperature
of the damper fluid.  This situation is inspired by a fluid damper where the
viscosity of the damper fluid, and hence the damping coefficient,
varies with temperature.  Here, the temperature of the fluid damper is
determined by an ODE that includes the effect of energy dissipated by
the damper and heat transferred from the damper to the surroundings,
which are assumed to have constant temperature.

The equations are as follows:
\begin{gather*}
m\ddot{x} + c(T) \dot{x} + kx = 0 \\
\dot{T} = c(T) \dot{x}^2 - \frac{1}{\tau}\left(T - T_0\right) \\
\end{gather*}
where $k$, $T_0$, and $\tau$ are constants and
\begin{equation*}
c(T) = \exp \left( \frac{T_0}{T} - 1 \right).
\end{equation*}
For all cases, the constants are set to the following values:
\begin{equation*}
k = 3, \quad T_0 = 20, \quad \tau = 1.
\end{equation*}
%



\end{document}